\documentclass[%
 aip,
 longbibliography,
rsi,%
 amsmath,amssymb,
 reprint,%
]{revtex4-1}

\usepackage{color}
\usepackage{graphicx}% Include figure files
\usepackage{dcolumn}% Align table columns on decimal point
\usepackage{bm}% bold math
  \usepackage{algorithm} %sagar
  \usepackage{algpseudocode}  %sagar
  \usepackage{multirow}
\usepackage{natbib}

\usepackage{tikz}
\usetikzlibrary{shapes.geometric, arrows}

\newcommand{\NNc}[1]{ {\color{red} \tt {#1}} }

\begin{document}

\preprint{AIP/123-QED}

%\title{{\it Ab Initio} MD Simulations with Hybrid Functionals: Implementation and Application}
\title{
Achieving an Order of Magnitude Speed-up in Hybrid Functional and Plane Wave based {\it {Ab Initio}} Molecular Dynamics: Applications to Proton Transfer Reactions in Enzymes and in Solution
%Efficient AIMD simulations with hybrid functionals using MTS scheme and ACE approach
%{\it Ab Initio} Molecular Dynamics Simulations of Chemical Reactions with Hybrid Functionals and Plane Wave Basis Set
}
%\title[Sample title]{Sample Title New:\\with Forced Linebreak\footnote{Error!}}% Force line breaks with \\
%\thanks{Footnote to title of article.}

\author{Sagarmoy Mandal}
\affiliation{ 
Department of Chemistry, Indian Institute of Technology Kanpur (IITK), Kanpur - 208016, India%\\This line break forced with \textbackslash\textbackslash
}%
\affiliation{ Interdisciplinary Center for Molecular Materials and Computer Chemistry
Center, Friedrich-Alexander-Universität Erlangen-Nürnberg (FAU),
Nägelsbachstr. 25, 91052 Erlangen, Germany
%\\This line break forced with \textbackslash\textbackslash
}%
\author{Vaishali Thakkur}
\affiliation{ 
Department of Chemistry, Indian Institute of Technology Kanpur  (IITK), Kanpur - 208016, India%\\This line break forced with \textbackslash\textbackslash
}%
\author{Nisanth N. Nair}
\email{nnair@iitk.ac.in}
 %\homepage{http://www.Second.institution.edu/~Charlie.Author.}
%\affiliation{%
%Second institution and/or address%\\This line break forced% with \\
%}%

\affiliation{ 
Department of Chemistry, Indian Institute of Technology Kanpur  (IITK), Kanpur - 208016, India%\\This line break forced with \textbackslash\textbackslash
}%

\date{\today}% It is always \today, today,

\begin{abstract}
{\it Ab initio} molecular dynamics (AIMD) with hybrid density functionals 
and plane wave basis
is computationally expensive 
due to the high computational cost of  
exact exchange energy evaluation.
% 
%Computational overhead can be substantially decreased by neglecting some of the 
% orbital products based on the extent of their contribution. % to exact exchange energy.
%
%For this, 
Recently, we proposed a strategy to combine adaptively compressed exchange (ACE) operator formulation and multiple time step (MTS) integration scheme to reduce the computational cost significantly [{\em  J. Chem. Phys.} {\bf 151}, 151102 (2019)]. 
%r-RESPA alorithm to increase the integration timestep. 
%
%Within AIMD, we achieve time scale separation in forces by introducing artificial splitting of the ionic forces based on adaptively compressed exchange (ACE) operator formulation.
%
%In this work, we presented the full theoretical details of  this method
%
%and demonstrated the computational efficiency through benchmarking the liquid water system. %
However, it was found that the construction of the ACE operator, which has to be done at least once in every MD time step, is %computationally expensive. 
% 
%Thus, construction of ACE operator remained as 
computationally expensive. 
In the present work, systematic improvements are introduced to further speed-up by employing localized orbitals for the construction of the ACE operator.
%
%Here, we present a new method which
%
%Specifically, the localized selected column of the density matrix (SCDM) orbitals are used
%approach to localize the orbitals in real space and use them 
%to build the ACE operator.
%
By this, we could achieve a computational speed-up of an order of magnitude for a periodic system containing 32-water molecules.
Benchmark calculations were carried out to show the accuracy and efficiency of the method in predicting the structural and dynamical properties of bulk water.
To demonstrate the applicability, computationally intensive free energy computations at the level of hybrid density functional theory were performed to investigate (a) methyl formate hydrolysis reaction in neutral aqueous medium and (b) proton transfer reaction within the active site residues of class-C $\beta$-lactamase enzyme.
%with hybrid functional based AIMD simulations.
%and scrutinized the accuracy of hybrid functional in predicting the free energy barrier.
%
%{\color {blue}
%Finally, we studied the protonation state of active site residues of class-C $\beta$-lactamase enzyme employing hybrid functional based
%molecular dynamics within the QM/MM framework.
%}
%
%Valid PACS numbers may be entered using the \verb+\pacs{#1}+ command.
\end{abstract}

\maketitle

\section{\label{sec:intro}Introduction}
{\it Ab initio } molecular dynamics (AIMD) techniques are extensively used to investigate  structural and dynamical properties of a wide variety of molecular systems, and to predict mechanism and free energetics of physiochemical processes.\cite{Tuckerman-book,leach,frenkel_smit,md_review_tuckerman} 
In AIMD simulations, inter-atomic interactions are evaluated on-the-fly using  electronic structure calculations at every MD step.\cite{marx-hutter-book,tuckerman_aimd,Tuckerman_2002_aimd,Payne_RevModPhys}
%is calculated through the electronic structure calculations.
%
%
 Kohn Sham density functional theory (KS-DFT) with plane wave (PW) basis set is widely used for performing AIMD simulations of periodic condensed matter systems.
%
%
%Plane wave (PW) basis sets are the preferred choice for KS-DFT calculations as they are  
%
%
%free from the basis set superposition errors and Pulay forces.
%
The choice of the exchange-correlation (XC) functionals for the KS-DFT calculations determines the accuracy of the predicted properties. 
%
%The accuracy of the KS-DFT based AIMD calculations critically depend on the chosen exchange-correlation (XC) functionals.
%
%
%For most of the applications, AIMD using PW KS-DFT is largely restricted to the
Generalized gradient approximation (GGA)\cite{PRA_GGA_Becke,PRB_GGA_LYP,PRL_GGA_PBE} type  XC functionals are most commonly used to carry out AIMD using PW KS-DFT.
However, these functionals suffer from self interaction error (SIE)\cite{Chemist's_Guide,Science_DFT_limitations,PRB_SIC,PRL_DFT_errors}
where the XC functional erroneously includes the unphysical self-interaction of electron density with itself.
%
%
%This error leads to over-delocalization of electron density, where 
Due to the SIE, the XC functionals tend to over-delocalize the electron density, leading to
%
%The delocalization error leads to
the underestimation of bandgap of solids, reaction barriers, and dissociation energies.\cite{Science_DFT_limitations,PRL_DFT_errors} 
%
%{\tt \color{red} WHAT ARE THE OTHER ERRORS? ONLY SIE?}

SIE can be minimized by using hybrid functionals, where a certain portion of the Hartree-Fock (HF) exchange energy is added to the GGA exchange energy.\cite{Chemist's_Guide,Martin-book,JCP_B3LYP,JCP_PBE0,JCP_HSE}
%
%{\tt \color{red} IS SIE THE ONLY REASON FOR DOING HYBRID XC?}
%
{%\color {red}
Hybrid functionals 
%partially reduce the self-interaction error and 
are generally known to improve the prediction of energetics, structures, electronic properties, chemical reaction barriers and band gap of solids.\cite{JCP_B3LYP,JCP_PBE0,JCP_PBE0_model,JCP_HSE,PCCP_HFX,PCCP_HSE,Galli_RSB_JPCL,Chem_Rev_Cohen,jpcl_2011_hfx,jpcl_2019_hfx,jpcl_2020_hfx,Scuseria:JCP:2008,Adamo:2012} 
%{\tt \color{red} HOPE YOU ARE NOT MISSING ANY RECENT PAPERS/REVIEWS?}
%
%Also, AIMD simulations with hybrid functionals 
%have been shown to
%provide an improved description of the structural and dynamical properties of liquids.\cite{JPCB_AIMD_HFX,JCTC_AIMD_HFX,JCP_AIMD_HFX,Mol_Phy_Car_MLWF,Mol_Phy_Car_MLWF_1,JPCB_water_hfx} 
%
Also, hybrid functional based AIMD simulations have been shown to improve the description of the structural and dynamical properties of liquids\cite{JPCB_AIMD_HFX,JCTC_AIMD_HFX,JCP_AIMD_HFX,Mol_Phy_Car_MLWF,Mol_Phy_Car_MLWF_1,JPCB_water_hfx} as well as
%in combination with enhanced sampling methods have been used
%are shown
%to improve 
the accuracy of the computed free energy surfaces.\cite{JCTC_Galli_FES,JCP_sagar,sagar_JCC}
}
However, AIMD simulation with hybrid functionals and PW basis set 
is extremely time consuming due to the high computational cost associated with the exact exchange energy evaluation.\cite{JCP_HFX_Voth} 
%
%
%{\color {blue} 
Thus, it is not a common practice to perform hybrid functional based AIMD simulations for systems containing several hundred or more atoms.
Various possible strategies have been proposed to speed-up such calculations, for example,
utilization of localized orbitals,\cite{PRB_Car_Wannier,JCP_AIMD_HFX,Mol_Phy_Car_MLWF,Mol_Phy_Car_MLWF_1,Nature_Car_MLWF,PRL_RSB,JCTC_RSB,JCTC_RSB_1,Galli_RSB_CPL,Galli_RSB_JPCL,Galli_RSB_JPCL1,JCP_sagar,Car_hfx_2019,enabling_part2} 
usage of the multiple time step (MTS) algorithms,\cite{HFX_Hutter_JCP,MTS_AIMD_Ursula,MTS_AIMD_Steele_3} 
%{use of single-precision arithmetic,}\cite{single_precision_hfx}
{use of coordinate-scaling approach,}\cite{JPCL_2018_Bircher,CPC_BIRCHER_2020}
employment of massively parallel algorithms\cite{HFX_Curioni,DUCHEMIN_2010,VARINI_2013,BARNES_2017} %TODO cite
{and other approaches}.\cite{jctc_Bolnykh_2019,single_precision_hfx}
Recently, we proposed a robust method\cite{JCP_2019_sagar,sagar_JCC} for performing efficient hybrid functionals and PW based AIMD, 
%using  in a computationally efficient way.
%
where a MTS integrator\cite{Tuckerman-book}
scheme was employed based on the adaptively compressed exchange (ACE)~\cite{ACE_Lin,ACE_Lin_1} operator 
formalism.
%
%an efficient splitting scheme to partition the ionic forces based on the adaptively compressed exchange (ACE) operator formulation.
%
In the proposed method, the ionic forces were artificially partitioned into computationally cheap fast forces and computationally costly slow forces with the help of an approximate representation of the exact exchange operator (i.e. the ACE operator).
Using the MTS algorithm, the computationally cheap fast forces were computed more frequently as compared to the computationally costly slow forces, 
thereby reducing the computational cost of performing such AIMD calculations.
%
%
%The implementation of our method in the CPMD code allows us to perform efficient hybrid functional based AIMD simulations with r-RESPA algorithm.
%
Additionally, our method has been shown to be efficient and accurate in predicting the structural and dynamical properties of realistic condensed matter systems.
It was realized that the construction of the ACE operator, which has to be done at least once in every MD time step, is %computationally expensive. 
% 
%Thus, construction of ACE operator remained as 
the computational bottleneck. 
In the present work, we systematically improve the %introduced to further 
computational speed by employing localized orbitals for the construction of the ACE operator.
%
%evaluation of the exact exchange integral.
%
%Here, 
%
% we present a strategy to combine the selected columns of the density matrix (SCDM) approach with the ACE operator based MTS scheme.
%
%Here, we present a new method which
%
Specifically, the localized selected column of the density matrix (SCDM)\cite{SCDM_main} orbitals are used
%approach to localize the orbitals in real space and use them 
to build the ACE operator. 
The SCDM method for obtaining localized orbitals is computationally 
efficient, and the procedure is not iterative (unlike 
in the case of computing Maximally Localized Wannier Functions).
Recently, Carnimeo {\it et al.}\cite{Carnimeo_2019} also employed a methodology to speed-up ACE operator construction based on the localized SCDM orbitals.
Their implementation in the {\tt Quantum ESPRESSO}\cite{Espresso_hfx} code reported a speed-up of at least 3-4 fold for hybrid functionals and PW based exact exchange energy calculations.
%with a reduced computational cost.
%efficiently.
%based localized orbitals to decrease the computational time for the construction of ACE operator.
%
%The combination of the MTACE scheme and SCDM orbitals based screening technique enables us to achieve further speed-up in the hybrid functional based AIMD calculations.
%
%Thus, it lowers the computational cost even futher and opens the door for simulatiing large sclae systems.
%
%This new technique is termed as s-MTACE.
%

Our approach presented in this paper, termed as s-MTACE, opens up the possibility of 
performing hybrid DFT based MD simulations of large condensed matter systems
providing a speed up of 10 fold or more.
%, opens up the possibility of 
%performing such AIMD simulations for large condensed matter systems.
%and intricate chemical reactions.
%
After presenting the theory and computational details, benchmark calculations are reported to demonstrate the accuracy and the computational efficiency of the method.
%with 32 water molecule system.
%
%Benchamark calculations were carried out with the 32 water molecule system with QM approach as well as with alanine dipeptide solvated in explicit water with the QM/MM approach.
%
%Accuracy of the proposed method was scrutinized by 
In particular, we have investigated the structural and dynamical properties of bulk water for benchmarking purpose.
%system from the s-MTACE and the conventional methods. 
%
%Also, the efficiency of the implemented method (in the {\tt CPMD} code) is examined for carrying out hybrid functional based AIMD simulations.
%with the proposed MTS algorithm.
%and allows us to gain one order of magnitude speed-up.
%
%Finally, we
Subsequently, we used our approach to perform free energy calculations of 
two important proton transfer reactions, namely methyl formate hydrolysis reaction in neutral water employing periodic  AIMD simulations (Figure~\ref{mech}), and proton transfer reaction within the active site residues of class-C $\beta$-lactamase enzyme complexed with cephalothin drug using quantum-mechanical/molecular-mechanical (QM/MM) simulations (Figure~\ref{prot_mech}).
Here we compare the free energies of chemical reactions computed using both PBE (GGA) and PBE0 (hybrid) density functionals.
Our results shed light on the performance of the PBE and 
the PBE0 functionals in predicting free energy barriers of reactions involving proton transfers in solutions and in enzymes
from finite temperature MD simulations.
%%%%%%%%%%%%%%%%%%%%%%%%%%%%%%%%%%%%%%%%%%%%%%%%%%%%%%
\section{\label{sec:theory}Theory}
%

%
%The self consistent field (SCF) solution of hybrid functional based 
KS-DFT calculations with hybrid functionals requires the application of the exact exchange operator ${\mathbf V}_{\rm X}=-\sum_{j}^{N_{\rm orb}} \frac{ | \psi_{j} \rangle  \langle \psi_{j} |}{r_{12}}$ on each of the KS orbitals $|\psi _{i} \rangle$ at each step of the self consistent field (SCF) iterations:
\begin{equation}
\label{vx}
\begin{split}
{\mathbf V}_{\rm X}|\psi _{i}\rangle & =- \sum_{j}^{N_{\rm orb}} |\psi _{j} \rangle \left \langle\psi _{j} \left | \left ( r_{12}\right )^{-1} \right | \psi _{i}\right \rangle ,  \\ & =- \sum_{j}^{N_{\rm orb}} v_{ij}(\mathbf{r}_{1}) |\psi _{j}\rangle \enspace, \enspace ~~i=1,....,N_{\rm orb}
\end{split}
\end{equation}
with
\begin{equation}  
\label{e:vij}
v_{ij}(\mathbf {r}_{1})=\left \langle\psi _{j} \left | \left ( r_{12}\right )^{-1} \right | \psi _{i}\right \rangle \enspace.   
\end{equation}
Here, $N_{\rm orb}$ is the total number of occupied orbitals.
Computation of $v_{ij}(\mathbf{r})$ is efficiently done in reciprocal space\cite{JCP_HFX_Voth,PRB_Car_Wannier} using Fourier transform.
%by taking advantage of the convolution theorem
%
%
If $N_{\rm G}$ is the total number of PWs, the computational cost for doing Fourier transform scales as $N_{\rm G}\log  N_{\rm G}$ by using the Fast Fourier transform (FFT) algorithm.
The total computational overhead scales as $N_{\rm orb}^2 N_{\rm G}\log  N_{\rm G}$,\cite{JCP_HFX_Voth} as the operation of ${\mathbf V}_{\rm X}$ on all the KS orbitals requires $N_{\rm orb}^{2}$ evaluations of $v_{ij}(\mathbf{r})$. 
As a result, such calculations are highly computationally intensive for typical molecular systems of our interest with about 100 atoms.

Whereas, the recently developed ACE operator formulation\cite{ACE_Lin,ACE_Lin_1} could greatly reduce the computational cost of such calculations.
In the ACE operator formulation, the full rank ${\mathbf V}_{\rm X}$ operator is approximated by the low rank ACE operator ${\mathbf V}_{\rm X}^{\rm ACE}= -\sum_{k}^{N_{\rm orb}}  | P_{k} \rangle  \langle P_{k} |$.
%using a low rank decomposition.
%
Here, $\{|P_{k} \rangle\}$ is the set of ACE projection vectors which can be computed by the decomposition of the ${\mathbf V}_{\rm X}$ operator; {See Appendix~A for more details.}
%easily through a series of linear algebra operations.
%, as explained below.
%
%
%The computation of $\{|P_{k} \rangle\}$ has to be performed 
Construction of $\{|P_{k} \rangle\}$ requires evaluation of $\{{\mathbf V}_{\rm X}|\psi _{i}\rangle \}$, which is a computationally costly step, because of the $N_{\rm orb}^{2}$ number of evaluations of $v_{ij}(\mathbf{r})$.\cite{ACE_Lin} 
However, once the ${\mathbf V}_{\rm X}^{\rm ACE}$ operator is constructed, the evaluation of the action of this operator on KS orbitals can be done easily with $N_{\rm orb}^{2}$ number of inner products as
\begin{equation}
{\mathbf V}_{\rm X}^{\rm ACE}|\psi _{i}\rangle=- \sum_{k}^{N_{\rm orb}} |P_{k} \rangle  \left \langle P_k | \psi_{i} \right \rangle , \enspace ~~i=1,....,N_{\rm orb}  \enspace .
\end{equation}
The advantage of the ACE approach is that the cost of applying the ${\mathbf V}_{\rm X}^{\rm ACE}$ operator on each KS orbitals is much less as compared to ${\mathbf V}_{\rm X}$ operator.
%

%
%According to this ACE scheme,
%at the first SCF step, ${\mathbf V}_{\rm X}^{\rm ACE}$ operator can be constructed through the computation of $\{{\mathbf V}_{\rm X}|\psi _{i}\rangle \}$, which is the costliest step (because of $N_{\rm orb}^{2}$ times evaluation of $v_{ij}(\mathbf{r})$).
%
%As HF exchange energy has only a minor contribution to the total energy, an approximate energy computation
%is possible by using the previously constructed ${\mathbf V}_{\rm X}^{\rm ACE}$ operator without updating it for the rest of the SCF iterations.
%
%Additionally, the ${\mathbf V}_{\rm X}^{\rm ACE}$ operator is updated with the newly obtained KS orbitals in an outer loop until a final convergence in HF energy is achieved.
%
%It is again stressed that, once the ${\mathbf V}_{\rm X}^{\rm ACE}$ operator is constructed, its low rank structure allows the easy computation of $\{{\mathbf V}_{\rm X}^{\rm ACE}|\psi _{i}\rangle \}$ in the subsequent SCF iterations.

%
In our earlier method,\cite{JCP_2019_sagar,sagar_JCC} we took advantage of this particular feature of the ACE operator to combine with the MTS scheme for performing efficient hybrid functional based AIMD. 
However, the construction of the $\mathbf V_{\rm X}^{\rm ACE}$ operator, which has to be done at least
once in every MD time step,
is computationally demanding.
%(and has the same computational 
%cost of applying the exact exchange operator). 
%
%Thus, in this method, construction of $\mathbf V_{\rm X}^{\rm ACE}$ operator remains as the computational bottleneck. 
%
%
From Equation~(\ref{e:vij}), it is clear that $v_{ij}(\textbf{r})$ will be zero for  
non-overlapping orbital pairs 
%in real space 
and such pairs of orbitals will not contribute to the sum in Equation~(\ref{vx}).
%to the HF exchange energy. 
%
%
Thus, it is possible to speed-up the construction of the ACE operator
%improve systematic improvements can be further made to  this approach, in particular
by employing a screened set of localized orbitals.\cite{Carnimeo_2019}
%this approach systematically
%improve systematic improvements can be further made to  this approach, in particular
%using a set of localized orbitals for the construction of the ACE operator.\cite{Carnimeo_2019}
%
%  
For systems with finite band gap, a unitary transformation can be carried out to localize the KS orbitals in real space as
\begin{equation} 
\label{unit}
|\phi _{k} \rangle=\sum_{i}^{N_{\rm orb}} |\psi _{i} \rangle u_{ik}  \enspace,
\end{equation}
where $u_{ik} \equiv \left ( \mathbf U \right )_{ik}$ and $\mathbf U$ is a unitary matrix.
%
%Both the ground state energy and the electron density are invariants under this unitary transformation.
%
Now, by a screening procedure, where least contributing orbitals are neglected, we expect to achieve a substantial decrease in the number of orbitals
involved in the evaluation of $\{{\mathbf V}_{\rm X}|\psi _{i}\rangle \}$.
%

%
%Now, the usage of this set of localized orbitals $\{|\phi _{k}\rangle \}$ in Equation~(\ref{vx}) allows a substantial decrease in the number of orbitals
%involved in the evaluation of $\{{\mathbf V}_{\rm X}|\psi _{i}\rangle \}$ through neglecting orbital
%pairs with insignificant contribution.
%, the computational cost now scales as
%
%$\overline N_{\rm orb}^2 N_{\rm G} \log N_{\rm G}$, where $\overline N_{\rm orb}^2$ 
%is the effective number of orbital products considered in the calculation
%after screening, and ideally $\overline N_{\rm orb}^2 \ll N_{\rm orb}^2$.
%

% 
In our present study, we employed the SCDM method\cite{SCDM_main} to localize the canonical KS orbitals in real space.
%
%It is a well known fact that each columns of the density matrix is spatially localized in case of systems with finite bandgap.
%
%In the SCDM method, $N_{\rm orb}$ number of linearly independent columns are selected from the density matrix via the rank revealing QR factorization.
%
%These selected columns serve as the localized orbitals (SCDMs) after an orthogonalization step.
%
%
%The algorithm for computing the HF exchange energy and the gradient of exchange energy with using SCDMs is given in Appendix \ref{scdm_algo}.
%
In our computations, we employed
%
%a pair density cutoff $\rho_{\rm cut}$ to screen the orbital pairs entering in  Eq.(\ref{vx}); an orbital pair $i$-$j$ is considered only if
%
the following cutoff criteria 
\[ \int d\textbf{r} \left | \phi_{i}(\textbf{r}) \phi_{j}^{*}(\textbf{r}) \right | \geqslant \rho_{\textrm {cut}} \enspace,\]
to screen the orbital pairs entering in Equation~(\ref{vx}).
An orbital pair $i$-$j$ is only considered during the computation of the $\mathbf V_{\rm X}^{\rm ACE}$ operator if the above criteria is satisfied.
%
%where $\rho_{ij}$, given by Eq.~(\ref{rho_ij}), is computed by the product of two localized orbitals $\phi_i$ and $\phi_j$.
%
%
The ACE operator computed through the above mentioned screening scheme is denoted as $\mathbf V_{\rm X}^{\textrm{s-ACE}}$ (screened ACE operator).

Now, similar to our earlier work,\cite{JCP_2019_sagar,sagar_JCC} 
%
%
%%%%%%%%%%%%%%%%%%%%%%%%%%%%%%%%%%%%%%%
%
we split %the contribution of ionic forces from the HFX part as
 the individual ionic force components (in hybrid functional based AIMD) for a system containing $N$ atoms as
\begin{equation}
    F^{\rm exact}_K= F^{\textrm{s-ACE}}_K+ \Delta F_K  \enspace , \enspace K=1,\cdots,3N    \enspace 
\end{equation}
with $\Delta F_K = \left ( F^{\rm exact}_K -  F^{\textrm{s-ACE}}_K \right )$.
Here, $\mathbf F^{\rm exact}$ and $\mathbf F^{\textrm{s-ACE}}$ are the ionic force computed with the full rank exchange operator ${\mathbf V}_{\rm X}$ and the low rank ${\mathbf V}_{\rm X}^{\textrm{s-ACE}}$ operator constructed (only) at the beginning of SCF cycles, respectively.
%(as shown in Fig.~\ref{slow_forces}).
%
%The term $\mathbf F^{\rm ACE}$ is the ionic force calculated using the low rank ${\mathbf V}_{\rm X}^{\rm ACE}$ operator (constructed at the beginning of SCF iteration cycles).
%as shown in Fig.~\ref{fast_forces}.
%
%
%
As the ${\mathbf V}_{\rm X}^{\textrm{s-ACE}}$ operator closely resembles the ${\mathbf V}_{\rm X}$ operator,
 the differences in the ionic force components of $\mathbf F^{\rm exact}$ and $\mathbf F^{\textrm{s-ACE}}$ are very small.
%and the two forces $\mathbf F^{\rm ACE}$ and $\Delta \mathbf F$ have different time scale for their variation.
%
%Thus, it is reasonable to
%
%employ the assumption that 
Now we make the assumption that
$\mathbf F^{\textrm{s-ACE}}$ is the fast force and $\Delta \mathbf F$ is the slow force. 
%
%
%In our earlier study in Chapter~\ref{chapter_2} (where ${\mathbf V}_{\rm X}^{\textrm{ACE}}$ operator was used instead of the ${\mathbf V}_{\rm X}^{\textrm{s-ACE}}$ operator), we have shown that this assumption is valid and these two forces vary with different timescales.
%
%
We have verified this assumption during our benchmark calculations.
Finally, we employ the reversible reference system propagator algorithm (r-RESPA)\cite{r-RESPA} to compute the computationally costly $\Delta \mathbf F$ forces less frequently as compared to the cheaper $\mathbf F^{\textrm{s-ACE}}$ forces, thereby speeding-up the calculations.
Specifically, $\mathbf F^{\textrm{s-ACE}}$ is computed at every small time step $\delta t$ and  $\Delta \mathbf F$ is computed at every $n$ steps (i.e., with a time step $\Delta t = n \delta t$).
Here, the small time step $\delta t$ can be chosen as per the time scale of fast forces ($\mathbf F^{\textrm{s-ACE}}$) that are cheaper to compute.
%
%NN: I am commenting this line.
%Whereas, the longer time step, $\Delta t$, can be decided according to the time scale of variation of the computationally costly slow forces ($\Delta \mathbf F$).
%, 
%
%In this way, we get the required speed-up to perform hybrid functional based AIMD simulations.
%
%Flowchart of the above mentioned scheme is shown in Fig.~\ref{mts}.
%
%
%Our description of r-RESPA till now allows us to carry out microcanonical (NVE) AIMD simulations.
%
%Here, it has to be noted that efficient thermostats have to be employed in order to eliminate the resonance effects originated with the use of large time step.\cite{mts_resonance,Resonance_MET} 
%

%\clearpage

\section{Computational Details}

All the calculations presented here were carried out employing { a modified version of}
% with its pseudopotential implementation~\cite{marx-hutter-book}
the {\tt CPMD} program\cite{cpmd,KLOFFEL2021} where the proposed s-MTACE method has been implemented.
%with our implemented method.
%
We used PBE0\cite{JCP_PBE0_model} and PBE\cite{PRL_GGA_PBE} functionals 
%as the hybrid XC functional 
for all the calculations.
The effect of the core electrons were incorporated using the 
norm-conserving Troullier-Martin type pseudopotentials,\cite{PRB_TM}
generated for PBE functional.
For expanding the wavefunctions in PW basis set, a cutoff energy of 80~Ry was used.
We carried out Born-Oppenheimer molecular dynamics (BOMD) simulations to perform MD simulations at microcanonical ($NVE$) and canonical ($NVT$) ensembles.
%simulations starting from a equilibrated NVT run\cite{JCP_sagar} with PBE0 functional.
%for all the systems.
%
%
%In order to perform canonical ensemble AIMD simulation, we employed
Nos{\'e}--Hoover chain thermostat\cite{NHC} was employed to perform canonical ensemble AIMD simulation at 300~K temperature.
The wavefunction convergence criteria was set to $10^{-6}$ a.u. for the orbital gradients during the SCF iterations.
%
%At every MD steps, wavefunctions were converged till the magnitude of maximum wavefunction gradient reached below $1\times 10^{-6}$ au.
%
For the initial guess of the wavefunctions at every MD step, Always Stable Predictor Corrector Extrapolation scheme\cite{JCC_ASPC} of order 5 was used.
\subsection{Benchmark Calculations: 32 Water System}
\begin{figure}
\includegraphics[scale=0.75]{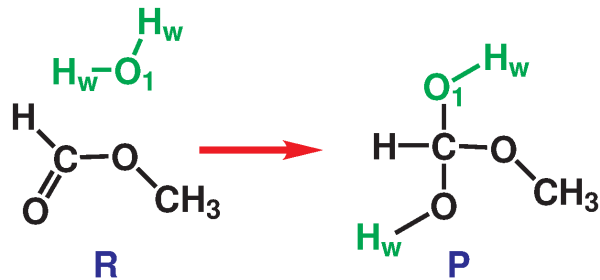}% Here is how to import EPS art
\caption{\label{mech} The formation of the gem-diol intermediate ({\bf P}) from the reactant ({\bf R})
during the methyl formate hydrolysis in neutral 
aqueous medium.
The attacking water molecule is shown in green color.
}
\end{figure}

\begin{figure}
\includegraphics[scale=0.48]{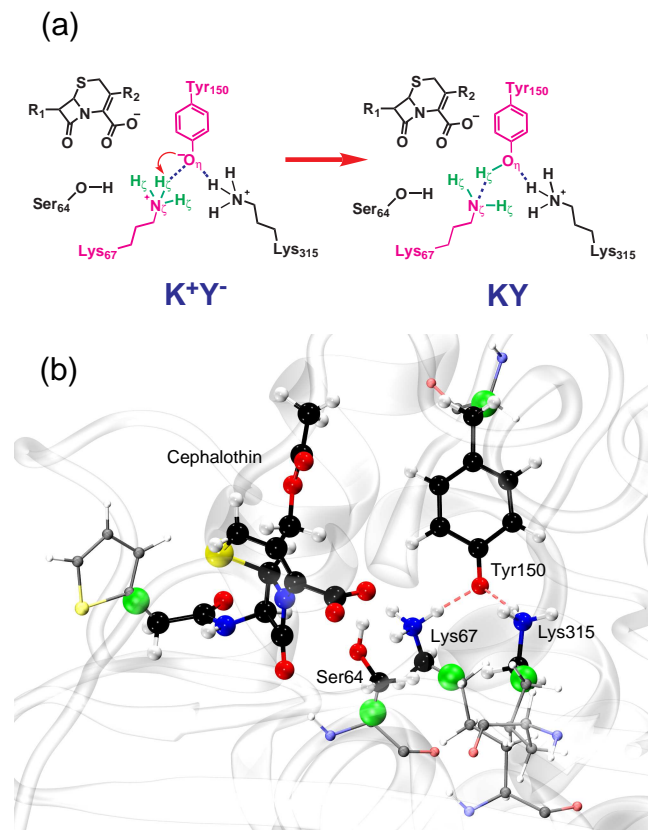}% Here is how to import EPS art
\caption{\label{prot_mech} (a) Proton transfer from protonated Lys$_{67}$ to Tyr$_{150}$ in presence of cephalothin ({\bf K$^+$Y$^-\rightarrow$} {\bf KY}). The residues involved in the proton transfer are shown in magenta color, and the hydrogen atoms involved are shown in green color.
(b) A representative snapshot of the solvated non-covalently complexed drug-substrate (class-C $\beta-$lactamase bound to the cephalothin drug molecule) in {\bf K$^+$Y$^-$} state.
Active site residues of class-C $\beta-$lactamase and the cephalothin drug molecule are shown in CPK style. 
The atoms shown in glossy spheres were treated by QM.
Atom color code: C (black), O (red), N (blue), S (yellow), H (white), and capping H (green).
%
%, and the hydrogen bonding interactions are shown using dashed lines in blue color.
%
%{\tt \color{red} MAKE FIGURE 3(b): ACTIVE SITE STRUCTURE WITH LABELS.}
}
\end{figure}

We carried out benchmark calculations for bulk water, modelled using 32 water molecules taken in a periodic supercell of dimensions 9.85~{\AA}$\times$9.85~{\AA}$\times$9.85~{\AA} corresponding to 
 water density $\sim$1~g~cm$^{-3}$.
Two classes of simulations were performed to benchmark the efficiency and the accuracy of our method:

(a) {\bf VV}: Here, we performed BOMD simulations in NVE/NVT ensemble with the standard velocity Verlet scheme with a timestep of $\Delta t$~fs. 
The value of $\Delta t$ used for the runs will be specified later in the
below sections.
% au. %($\approx 0.5$~fs).
%
%This run was performed to generate 10 ps long trajectory.

(b) {\bf MTS-n}: Here, we performed BOMD simulations in NVE/NVT ensemble with s-MTACE scheme with an inner time step, $\delta t =  0.5$~fs and an outer time step, $\Delta t = 0.5\times n$ fs.
%
%All these simulations were carried out to generate 15 ps long trajectories.
%TODO

\begin{figure}
\includegraphics[scale=0.12]{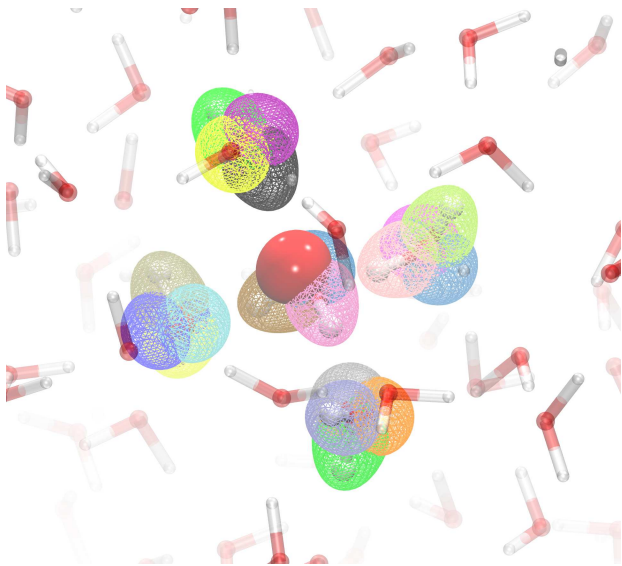}% Here is how to import EPS art
\caption{\label{scdm_fig}
%The overlap of a particular SCDM orbital (shown with red solid surface) with the nearest neighbour SCDM orbitals (shown in wireframe representation with different colors) for a periodic system of 64 water molecules.
%
%SCDM orbitals are shown for a central water molecule and the water molecules that are H-bonded to it.
%
%The SCDM with solid red color surface overlaps with all the other SCDMs shown in isosurface representation.
%
%Color code: red for O atoms and white for H atoms.
%An isovalue of 0.01 a.u. is used to obtain the contour surfaces of the SCDMs.
%
%64 water 0.01 isovalue
%
The SCDM-localized orbitals $\{ \phi_j \}$ (wireframe volumetric representation) considered in the computation of $v_{ij}$ for a given SCDM-localized orbital $\phi_i$ (solid-red volumetric representation) 
while using the cutoff $\rho_{\rm cut}=2.5 \times 10^{-2}$ for a periodic system containing 64 water molecules.  Atom color code: Red (O), White (H).
Isovalue used for the volumetric representation is $1\times 10^{-2}$ a.u.
}
\end{figure}

%
Additionally, for these {\bf MTS-n} runs, three different cutoffs ($\rho_{\rm cut}$) were chosen for constructing $\mathbf V_{\rm X}^{\textrm{s-ACE}}$ operators: 
 $\rho_{\rm cut}=2.0\times10^{-3},~1.0\times10^{-2}, {\textrm { and }}  2.5\times10^{-2}$. %
For obtaining a realistic picture, we mention in passing that while using $\rho_{\rm cut}=2.5\times10^{-2}$, 
Equation~\ref{e:vij} is mostly computed over the localized
orbitals within the first solvation shell (Figure~\ref{scdm_fig}).
%we consider only the overlap of a localized orbital on a particular water molecule with the other localized orbitals localized on water molecules that are present within the first solvation shell, as shown in Figure~\ref{scdm_fig}.
%

\subsection{Application: Methyl Formate Hydrolysis}
A cubic periodic simulation cell with side length of 10.1~{\AA} was chosen for modelling methyl formate hydrolysis in neutral water. The system contained one methyl formate molecule and 32 water molecules.
%
%All the calculations were carried out employing
% with its pseudopotential implementation~\cite{marx-hutter-book}
%the {\tt CPMD} program\cite{cpmd} wherein the r-RESPA+s-ACE method was  implemented by us.
%with our implemented method.
%
We performed two sets of simulations with the GGA/PBE and hybrid/PBE0 XC functionals.
%were used together with the norm-conserving Troullier-Martin type pseudopotentials\cite{PRB_TM}
%(constructed for the PBE functional\cite{PRL_GGA_PBE}) 
%a%nd a PW cutoff energy of 80~Ry was taken.
%was used.
%
BOMD simulations were carried out to perform MD simulations in canonical ensemble at $T=300$K temperature employing Nos\'e-Hoover chain thermostats.\cite{NHC} 
%{\tt \color{red} Mention about the thermostat}
%
For the PBE calculations, the standard velocity Verlet scheme was employed with a timestep of $\Delta t = 0.48$~fs.
Whereas, for the PBE0 runs, s-MTACE with $\delta t = 0.48$~fs and  $\Delta t = 7.2$~fs (i.e., $n=15$) were performed.
We used $\rho_{\rm cut}=2.5\times10^{-2}$ for the screening of the SCDM orbital pairs.

Well-sliced metadynamics (WS-MTD) approach\cite{JCC_shalini} was employed to compute the free energy surface of the methyl formate hydrolysis reaction.
%in neutral water.
%
%For this particular reaction, the WS-MTD method allows us to achieve a controlled sampling on the distance between a reactive water molecule and the methyl formate.
%of coordinates and efficient exploration of high-dimensional
%free energy landscape.
%
%This method is ideal for investigating the methyl formate hydrolysis reaction, 
%since we would like to sample the distance between a reactive water molecule and the methyl formate in a controlled manner, while ensuring exhaustive sampling of 
%the protonation state of the attacking water.
%
Two collective variables (CVs), ${\mathbf s =  \{ s_1, s_2 \} }$, were chosen to explore the free energy surface of the reaction.
%
%For the well-sliced metadynamcis simulation of the hydrolysis reaction we chose two CVs.
%
%The first CV ($s_1$) is 
The distance between the carbonyl carbon atom (C) of the methyl formate and the oxygen atom (O$_1$) of the attacking water molecule ($d$[C--O$_1$]) is chosen as the first CV ($s_1$), see Figure~\ref{mech}.
Sampling along this CV was performed in a controlled manner using the umbrella sampling\cite{US_method} like 
bias potential
% 
%$s_1$:
\begin{eqnarray} 
\label{e:us:bias}
W_h(s_1)=\frac{1}{2}\kappa_h \left ( s_1 - d_h^{(0)} \right )^2 , ~~h=1,...,M \enspace .
\end{eqnarray}
%
%which ensures a controlled sampling along this coordinate.
%
Here, $M$ is the total number of umbrella windows used.
%while
%
$\kappa_h$ and $d^{(0)}_h$ are the restraining force constant and the equilibrium value of the $h$-th umbrella restraint, respectively.

%
%The  is 
The coordination number ($CN$) of the oxygen atom (O$_1$) of the attacking water with all the hydrogen atoms (H$_{\textrm w}$) of the solvent water molecules is chosen as the second CV ($s_2$) 
\begin{equation}
\label{CN}
CN[{\mathrm {O_1:H_w}}]=\sum_{i=1}^{N_{\mathrm {H_w}}} \frac{ 1}{1+({d_{1i}}/{d_0})^{6}} \enspace ,
\end{equation}
with $d_0=1.30$~{\AA}.
Here, $N_{\mathrm {H_w}}$ is the total number of H$_{\textrm w}$ atoms and $d_{1i}$ is the distance between the O$_1$ atom and the $i$-th H$_{\textrm w}$ atom.
%The value of the parameter .
%
This CV was sampled employing the well tempered metadynamics (WT-MTD) bias potential\cite{WT-MTD} 
%in Eq.~(\ref{e:mtd:bias}).
%
%and $V^{\rm b}(s_2,t)$ is the well tempered metadynamics bias
%potential\cite{WT-MTD} acting (only) on $s_2$:
\begin{eqnarray} 
\label{e:mtd:bias}
V^{\rm b}(s_2,t) = \sum_{\tau < t} w(\tau) \exp \left [ -\frac{ \left \{ s_2 - s_2(\tau) \right \}^2 }{2 \left ( \delta s \right )^2 } \right ] \enspace. \end{eqnarray}
Here, $\delta s$ is the width of the Gaussian function and the height of the Gaussian, 
$w(\tau)$, is given by,
\begin{equation}
 w(\tau) = w_0 \exp \left [ - \frac{ V^{\rm b}(s_2,\tau) }{k_{\rm B} \Delta T} \right ] \enspace,
 \end{equation}
where $w_0$ and $\Delta T$ are parameters and $k_{\rm B}$ is the Boltzmann constant.

{
For each umbrella window $h$, we construct the partially reweighted probability distribution 
\begin{equation}
\label{reweight}
P_h( s_{1}^\prime, s_{2}^\prime ) = \frac{\int_{t_{\rm min}}^{t_{\rm max}} dt \,  A_h(t)  \prod_{\alpha=1}^2 \delta \left ( s_\alpha(t) - s_\alpha^\prime \right ) }{ \int_{t_{\rm min}}^{t_{\rm max}} dt \,  A_h(t) }   
\end{equation}
with 
\begin{equation} 
A_h(t) = \exp \left [  \beta  \left \{ V_h^{\rm b}(s_2(t), t) - c_h(t) \right \} \right ] \nonumber
\end{equation}
and 
\begin{equation}
c_h(t) = \frac{1}{ \beta} \, \ln \left [  
\frac{\int ds_2 \exp \left [ \beta  \, \theta \, V^{\rm b}(s_2,t) \right ]  }
{\int ds_2 \, \exp \left [  \beta \left ( \theta - 1 \right ) \, V^{\rm b}(s_2,t) \right ]}
\right ] \enspace . \nonumber
\end{equation}
Here, $\theta=\left ( T + \Delta T \right )/\Delta T$, and $\beta=1/{ k_{\rm B}T}$.
%
%Here, the subscript $h$ runs over all $M$ umbrella windows.
%
The integrals in Equation~(\ref{reweight}) were evaluated for a time series from $t_{\rm min}$ to $t_{\rm max}$. 
Finally, the weighted histogram analysis method (WHAM)\cite{WHAM_JCC} was used to combine the partially reweighted probability distributions $\{P_h(s_{1}, s_{2}) \}$ to get the
Boltzmann reweighted probability distribution and free energy.
}

We have used 36  windows in total for sampling $d$[C--O$_1$] in the range from 1.51 to 3.70~{\AA}. 
%during the PBE0 (PBE) based simulations.
%
The parameters of the umbrella sampling bias potential ($\kappa_h$ and $d_h^{(0)}$) are
reported in Appendix~B1.
%taken according to our earlier study.\cite{JCP_sagar}  %TODO CITE
%shown in Appendix \ref{umbrella_details}. 
%
The time-dependent bias potential, $V^{\rm b}(s_2,t)$, acting along $CN[{\mathrm {O_1:H_w}}]$
was updated every 19.4~fs.
%during the production runs.
%
The parameters for the WT-MTD bias potential: $w_0$, $\delta s$ and $\Delta T$ were chosen to be 0.59~kcal~mol$^{-1}$, 0.05, and 4000~K, respectively. 
Initially, we carried out 2--3~ps of equilibration for each umbrella window.
%without adding the WT-MTD bias. 
%
%Then, 
Afterwards, 35~ps of production runs were performed for every window. 
Additionally, to sample the transition state region more extensively, we generated 46 and 50~ps long trajectories for
the windows near the transition state region ($s_1 \in [1.70, \, 2.05]$~{\AA})
for PBE0 and PBE based simulations, respectively.
%

%%%%%%%%%%%%%%%%%%%%%%%%%%%%%%%%%%%%%%%%%%%%%%%%%%%%%%%%%%%%%%%%%%%%%%

{

\subsection{Application: Protonation State of Active Site Residues of Class-C $\beta$-Lactamase}
The equilibrated initial structure for the solvated non-covalently complexed drug-substrate (class-C $\beta-$lactamase bound to the cephalothin drug molecule) was taken from an earlier work by our group;\cite{Ravi_2012_JPCB} { See Figure~\ref{prot_mech}(b)}.
%TODO: INCLUDE A ZOOMED FIGURE 
%OF THE ACTIVE SITE WITH PROTEIN AT THE BACKGROUND. YOU CAN COMBINE THIS FIGURE WITH FIGURE 3
%
The protein-drug complex was solvated with 13473 TIP3P water molecules in a periodic box of size $78\times77\times76$ \AA$^3$.
We used the CPMD/GROMOS interface as available in the {\tt CPMD} program\cite{cpmd} to carry out the hybrid QM/MM canonical ensemble MD simulations.
The protein side chains of Lys$_{67}$, Tyr$_{150}$, Lys$_{315}$, Ser$_{64}$, and the cephalothin molecule {(except the thiophene ring)} were treated by QM, and the remaining part of the system including the solvent molecules were modelled by MM.
Capping hydrogen atoms were used whenever the QM/MM boundary cleaves a chemical bond.
%
%The capping was introduced between C$_\delta$ and C$_\epsilon$ for Lysine residues, and between C$_\alpha$ and C$_\beta$ for Serine and Tyrosine residues.
%
In total 66 atoms were taken in a QM box with dimensions $18\times21\times22$ \AA$^3$; {see Figure~\ref{prot_mech}(b) for details}. %TODO include a figure with QM part in CPK format.
The QM part was treated using KS-DFT and PW.
{The core electrons of all the QM atoms were accounted using norm-conserving Troullier-Martin type pseudopotentials at the level of PBE and PW cutoff of 70~Ry was taken.\cite{PRB_TM}
MM part was treated using { parm99 AMBER}\cite{parm99} force-field.
%
%Ultrasoft pseudopotentials\cite{Ultrasoft:PP}, 
The QM/MM interaction was handled using the electronic coupling scheme developed 
by Laio {\it et al}.\cite{Coupl:Laio}
The QM charge density interacts explicitly with any MM point charge within a range of 15 \AA, beyond which the interaction was accounted by considering only the 
multipole expansion of the charge density up to the quadrapole term.
Nearest neighbor lists were updated every 50~steps, and the long-range 
interaction cutoff of 20~{\AA} was taken.
QM/MM BOMD simulations were carried out and the temperature of the system was maintained at 300K using two separate Nos\'e-Hoover chain thermostats\cite{NHC} for the QM and MM subsystems.}
%
%
%The QM part of the system was treated with PW KS-DFT with norm-conserving pseudopotentials and a PW cutoff of 70 Ry.
%
%
%

We performed two sets of simulations: 
(a) {\bf PBE}: the standard velocity Verlet scheme was employed with a timestep of $\Delta t = 0.48$~fs and the PBE XC functional was used.
(b) {\bf PBE0}: the s-MTACE scheme with $\delta t = 0.48$~fs and  $\Delta t = 7.2$~fs (i.e., $n=15$) were considered and the hybrid PBE0 XC functionals was used.
We took $\rho_{\rm cut}=2.0\times10^{-3}$ for screening the SCDM-localized orbital pairs.

Umbrella sampling approach\cite{US_method} was employed to model the
proton transfer reaction within the active site residues  of class-C $\beta$-lactamase enzyme.
In particular, we compute here the free energy barrier for the proton transfer between the Lys$_{67}$N$_{\zeta}$ and Tyr$_{150}$O$_{\eta}$ atoms of class-C $\beta$-lactamase in the presence of the substrate; see Figure~\ref{prot_mech}(a). 
The CN of the Tyr$_{150}$O$_{\eta}$ atom with all the three Lys$_{67}$H$_{\zeta}$ atoms ($CN$[Tyr$_{150}$O$_{\eta}$:Lys$_{67}$H$_{\zeta}$]) was biased using the umbrella potential as in  Equation~(\ref{e:us:bias}).
The definition of the CN is similar to Equation~(\ref{CN}).
A total of 24 windows were placed in the range 0.10 to 0.90.
%for sampling $CN$[Tyr$_{150}$O$_{\eta}$:Lys$_{67}$H$_{\zeta}$].
%
Structures with CN lying between 0.0 and 0.5 resemble $\mathbf{K^+Y^-}$, while those between 0.5 and 1.0 
are similar to $\mathbf{KY}$ state; {See Figure~\ref{prot_mech}(a)}.
The details of the umbrella bias potentials are reported in Appendix~B2.
{We generated 24 (35)~ps long trajectories per
each umbrella window during PBE0 (PBE) based QM/MM simulations.}
The biased probability distributions obtained from these independent simulations were then combined using the WHAM\cite{WHAM_JCC} technique to get the Boltzmann reweighted probability distribution, and hence the free energy.
}

\section{\label{sec:result_diss}Results and Discussion}
\subsection{\label{bench_calc} Benchmark Calculations}
%%%%%%%%%%%%%%%%%%%%%%%%%%%%%%%%%%%%%%%%%%%%%%%%%%%%%%%%%%%

\begin{figure}
\includegraphics[scale=0.32]{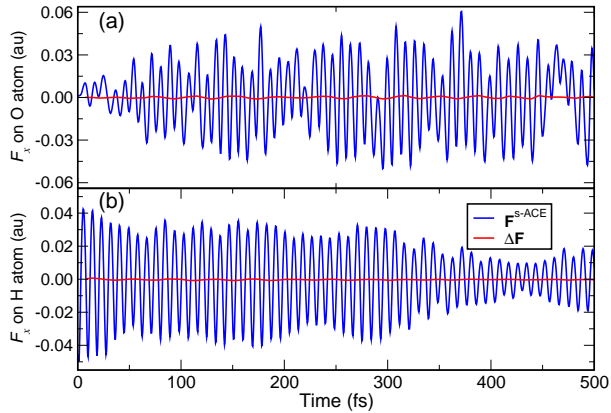}% Here is how to import EPS art
\caption{\label{f_component}
%Test results for 32-water system using PBE0 functional:
One of the components of $\mathbf F^{\textrm{s-ACE}}$ and $\Delta \mathbf F$ on an arbitrarily chosen (a) oxygen and (b) hydrogen atoms for a system containing 32 water molecules in a periodic box during AIMD simulation using PBE0 functional.
Screening of the SCDMs-localized orbitals was performed with $\rho_{\rm cut}=2.5\times10^{-2}$.
%
%(c)  Comparison of potential energy during {\bf VV}, {\bf MTS-5} and {\bf MTS-15} simulations in NVE ensemble; 
%
%All simulations were started from the same initial coordinates and velocities.
%
%(d)  $\log_{10}(\Delta E)$ for different $\Delta t$ values in {\bf VV} and {\bf MTS} simulations calculated from 5~ps long trajectories. %NN1: TODO: take log_10 and not log_e.
%
}
\end{figure}

\begin{figure}
\includegraphics[scale=0.7]{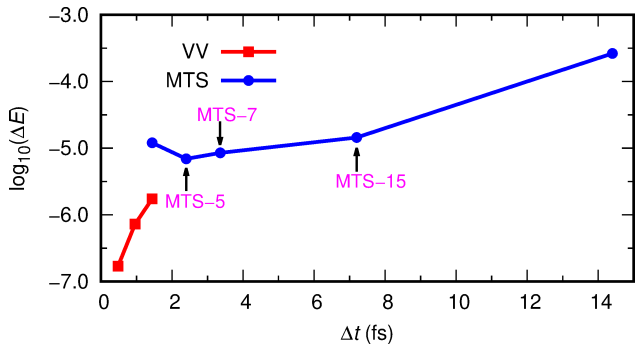}% Here is how to import EPS art
\caption{\label{fluc_ener}
%Test results for 32-water system using PBE0 functional:
%One of the components of $\mathbf F^{\rm ACE}$ and $\Delta \mathbf F$ on an arbitrarily chosen (a) oxygen and (b) hydrogen atoms.
%
%All simulations were started from the same initial coordinates and velocities.
%
%(d)  
Measure of the fluctuations in total energy, $\log_{10}(\Delta E)$, for different $\Delta t$ values in {\bf VV} and {\bf MTS-n} simulations using PBE0 functional.
%calculated from 5~ps long trajectories. %NN1: TODO: take log_10 and not log_e.
%
$\Delta E$ is calculated using Equation~(\ref{energy_con}) over 5~ps long trajectories.
In these runs, the screening of the SCDM-localized orbitals was performed with $\rho_{\rm cut}=2.5\times10^{-2}$.
}
\end{figure}
%

%\clearpage

\begin{figure}
\includegraphics[scale=0.45]{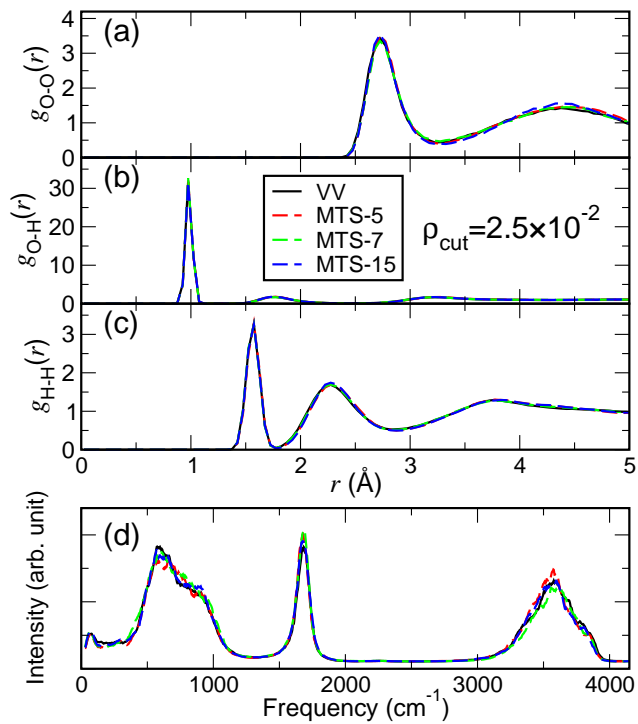}
\caption{\label{gofr_high}
Radial distribution functions (RDFs) from {\bf VV}, {\bf MTS-5}, {\bf MTS-7} and {\bf MTS-15} 
simulations of bulk water using PBE0 functional: (a) O-O, (b) O-H,  and (c) H-H.
(d) Power spectrum of the system computed from these simulations is also presented.
%
%Screening of the SCDMs was performed with $\rho_{\rm cut}=2.5\times10^{-2}$.
%
}
\end{figure}

%

%%%%%%%%%%%%%%%%%%%%%%%%%%%%%%%%5

%\clearpage

%%%%%%%%%%%%%%%%%%%%%%%%%%%%%%%%%%%%%%%%%%%%%%%%%%%%%%%%%%%%%%%%%%%
To test the accuracy and efficiency of the s-MTACE method, 
%of the timescale separation imposed, we evaluated $\mathbf F^{\textrm{s-ACE}}$ and $\Delta \mathbf F$ force components along a trajectory of 
we considered a periodic system containing 32 water molecules modelling bulk water.
%in a periodic box. 
%
In Figures~\ref{f_component}(a) and (b), we have shown one  component of $\mathbf F^{\textrm{s-ACE}}$ and $\Delta \mathbf F$ (calculated every $n=\Delta t/\delta t$ steps) on an arbitrarily chosen oxygen and hydrogen atom, respectively.
%
%where
%$\mathbf V_{\rm X}^{\rm ACE}$ is calculated once at the beginning of a SCF while kept fixed during the remaining SCF cycles, and 
%Here, $\Delta \mathbf F$ is calculated every $n=\Delta t/\delta t$ steps.
%
It is apparent that $\Delta \mathbf F$ is slowly varying as compared to $\mathbf F^{\textrm{s-ACE}}$.
Additionally, the magnitude of $\Delta \mathbf F$ is $\sim$100 times smaller as compared to
$\mathbf F^{\textrm{s-ACE}}$.
Thus we conclude that the time scale separation implied in the s-MTACE method is reasonable, as seen
in the case of MTACE method.\cite{JCP_2019_sagar}
%proposed in this work is applicable.
%

%As discussed in our previous work,\cite{JCP_2019_sagar} 
It is apparent that the efficiency of the MTS schemes crucially depends on the choice of the parameter $n$.
%outer timestep $\Delta t=n \, \delta t$.  
%
%A choice of larger $n$ will aid us in achieving better speed-up, as the computationally costly $\Delta \mathbf F$ forces have
%to be evaluated less frequently. 
%
%However, usage of a large $n$ value can degrade the quality of the trajectories associated with high drift in total energy.
%integration of the equations of motion and could result in total energy drift.
%
%The choice of $n$ is restricted by 
%The appearance of the resonance effects\cite{mts_resonance,Resonance_MET,TamarSchlick:Book} also restricts the use of the larger outer timestep $\Delta t$.
%which arises when the update frequency of the slow forces is proportional to its natural frequency.
%of the system.
%
%
In practise, one has to determine an optimal value of $n$ by monitoring the drift in total energy.
%in this proposed s-MTACE scheme.
%
%they are accompanied by larger drift in total energy.
%Based on those results, we choose $n_t=15$ in this work.\cite{JCP_2019_sagar}
%
%
% 
For this purpose, we compared the total energy fluctuations 
in velocity Verlet ({\bf VV}) and MTS runs ({\bf MTS-n})
with different $\Delta t$ for 32-water system in $NVE$ ensemble.
%, treated with PBE0 functional.
%
To measure the magnitude of the total energy ($E$) fluctuations, we computed the quantity %$\Delta E$ using Equation~\ref{energy_con} 
\begin{equation}
\label{energy_con}
\Delta E= \left < \left | \frac{E-\left \langle E \right \rangle}{\left \langle E \right \rangle} \right |  \right >   \enspace ,
\end{equation}
for various values of $n$.
%
%where $\left < \cdots \right >$ denotes time average.
%\begin{equation}
%\label{energy_con3}
%
%\Delta E= \left < \left | \frac{E-\left \langle E \right \rangle}{\left \langle E \right \rangle} \right |  \right >   \enspace ,
%\end{equation}
%
%where $\left < \cdots \right >$ denotes time average.
%
% 
%
In Figure~\ref{fluc_ener}, $\log_{10}(\Delta E)$ is plotted as a function of the timestep $\Delta t$ for {\bf VV} and {\bf MTS-n} runs, similar to the analysis done in the earlier works.\cite{MTS_AIMD_Ursula,JCP_2019_sagar}
In the case of {\bf VV} runs, it is clear that the total energy fluctuations increase with $\Delta t$.
%
%
%
%It has also been observed that the usage of timestep greater than 1.4~fs leads to unstable and erroneous trajectories in {\bf VV} runs.
%
%We also observed that the use of a timestep greater than 1.4~fs 
%in {\bf VV} runs leads to unstable trajectories with breaking of O-H covalent bonds.
%
For the {\bf MTS-n} runs, the inner timestep $\delta t$ is kept fixed at 
$\sim$0.5~fs 
and the outer timestep $\Delta t=n \, \delta t$ was varied.
%
%
%The quality of the energy conservation in these runs depends on the value of $n$, which determines how large the outer timestep is compared to the inner timestep.
%
In Figure~\ref{fluc_ener}, the slope of the curve is smaller for the {\bf MTS-n} runs as compared to the {\bf VV} runs, 
{indicating that the effect of increasing timestep is less profound in the MTS runs.}
%
%Also, the total energy conservation in
%{\bf MTS-n} runs with $n=15$ is comparable to {\bf VV} run using a timestep of $1.4$~fs {\tt \color{red} $\Rightarrow$ I THINK THIS STATEMENT IS NOT TRUE??}.
 %
The stability of the {\bf MTS-n} runs with $n=15$ (i.e. {\bf MTS-15}) demonstrates that it is a good choice for production runs. 
%for 32-water periodic system.
%
%$\Delta t=14.4$~fs) 
Whereas, {\bf MTS-n} simulations with $n$ greater than 15 were showing substantially high total energy fluctuations and therefore, they were not considered further.
%it was able 
%to generate stable MD trajectories.
%
Same conclusions about the optimal $n$ value were obtained while using
%
%A similar trend in the total energy fluctuations is also observed with 
the different $\rho_{\rm cut}$ values (see Appendix~C).
%
%Also, the drift in total energy for various {\bf VV} and {\bf MTS-n$_t$} simulations (with different $\rho_{\rm cut}$ values) are reported in Table~\ref{table_drift_nve}.

To benchmark the accuracy, stability, and efficiency of our proposed method, we performed a few sets of $NVT$ simulations with 32-water bulk system.
The details of the simulation length, average temperature and drift in total energy for these {\bf VV} and {\bf MTS-n} simulations are reported in Appendix~D.
The accuracy of the method is benchmarked by comparing the structural and dynamical properties of bulk  water obtained from {\bf VV} (which invokes no assumption) and {\bf MTS-n} simulations.
In particular, for comparing the structural properties,  partial radial distribution functions (RDFs) 
were computed; See Figure~\ref{gofr_high}(a)-(c).
It is clear that the location of the peaks and the peak heights of the
RDFs from the {\bf MTS-n} simulations are in excellent agreement with those from the {\bf VV} run.
For comparing the dynamical properties, we computed the power spectrum for the same system (shown in Figure~\ref{gofr_high}(d)) by taking the Fourier transform of the velocity auto-correlation function.
We have observed that the frequencies and the intensities of the spectra from {\bf VV} and {\bf MTS-n} 
simulations are in excellent agreement with each other.
These results demonstrate that the s-MTACE scheme provides an accurate description of structural and dynamical properties.
We have observed a similar trend for different values of $\rho_{\rm cut}$; See Appendix~E.
%{Section V} of the Supporting Information. 

%
Now, we study the computational efficiency of our method for which we have compared the average computational time {for generating 1~fs trajectory} with various parameter settings on an identical
computing platform.
The CPU time and the achieved speed-ups for various {\bf MTS-n} runs as compared to the {\bf VV} run are reported in Table~\ref{table_32} for periodic systems with 32, 64 and 128 water molecules.
Here, we defined the speed-up for a {\bf {MTS-n}} run as the ratio between the CPU time per {fs} in {\bf {VV}} ($t_{\mathbf {VV}}$) and {\bf {MTS-n}} runs ($t_{{\textbf {MTS-n}}}$):
\begin{equation}
    {\textrm {speed-up}}= \frac{t_{\mathbf {VV}}}{t_{\textbf {MTS-n}}} \enspace .
\end{equation}
%
%Here, we can calculate t
The time taken per {fs} in a {\bf {MTS-n}} run is calculated as:
\begin{equation}
 t_{\textbf {MTS-n}}=\frac{t_{\rm exact}^{\rm force}+n~t_{\textrm{s-ACE}}^{\rm force}}{n} { 
  \left ( \frac{\mathrm{1~fs}}{\delta t \, \, \mathrm{fs}} \right ) }  \enspace .
\end{equation}
Here, $t_{\rm exact}^{\rm force}$ and $t_{\textrm{s-ACE}}^{\rm force}$ are the times taken for $\mathbf F^{\rm exact}$ and $\mathbf F^{\textrm{s-ACE}}$ force calculations, respectively.
{
In every $n$ MD steps, the force using the exact exchange operator (i.e. $\mathbf F^{\rm exact}$) is calculated only once, and the force using the approximate s-ACE operator (i.e. $\mathbf F^{\textrm {s-ACE}}$) is computed $n$ times. 
The total CPU time for the computation of the $\mathbf F^{\rm exact}$ and $\mathbf F^{\textrm{s-ACE}}$ forces are decomposed into various contributions in Table~\ref{table1}. }
%
%The CPU time requirements for the computation of the different parts of these forces are reported in Table~\ref{table1}.
%

\begin{figure}[h]
\centering\includegraphics[scale=0.7]{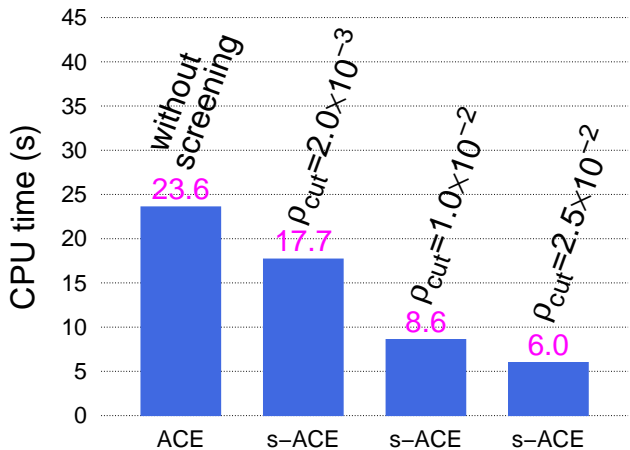}
\caption{\label{cost_ace}
%Test results for 32-water system using PBE0 functional:
%
%
Average computing time for the construction of 
${\mathbf V}_{\rm X}^{\textrm{s-ACE}}$  
for various values of $\rho_{\rm cut}$
for a periodic system containing 32 water molecules and
using 120 identical processors;
%per MD step for PBE0 calculation
%of a system containing 32 water molecules in a periodic box
%using various methods:
%
%Using the exact $\mathbf V_{\rm X}$ operator ({\bf VV}); using MTACE with $n_t = 15$ ({\bf MTS-15}) and using s-MTACE with $n_t = 15$ ({\bf MTS-15}) for various values of $\rho_{\rm cut}$.
%
%
%All the computations were using identical 120 processors.
%
%The computational time reported here is averaged over 630 MD steps;
%
%The achieved speed-ups for various {\bf MTS-15} runs compared to {\bf VV}, are shown in red.   
%
see also Table~\ref{table1}.
}
\end{figure}

\begin{figure}[h]
\centering\includegraphics[scale=0.7]{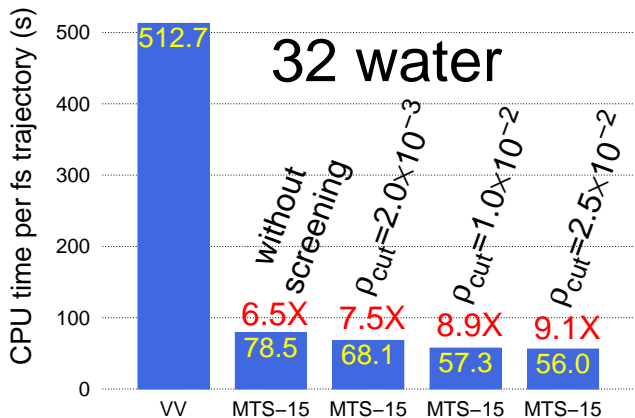}
\caption{\label{scaling}
%Test results for 32-water system using PBE0 functional:
%
%
Average computational time for generating 1~fs trajectory with PBE0 
for a system containing 32 water molecules in a periodic box
using the exact $\mathbf V_{\rm X}$ operator ({\bf VV}),
%using MTACE with $n = 15$ ({\bf MTS-15}) 
and s-MTACE with $n = 15$ ({\bf MTS-15}) for various values of $\rho_{\rm cut}$.
All the computations were performed using identical 120 processors.
The computational time reported here is averaged over 630 MD steps.
%
%The achieved speed-ups for various {\bf MTS-15} runs compared to {\bf VV}, are shown in red.   
%
%see also Table~\ref{table1}.
%
}
\end{figure}

From Table~\ref{table1}, it is clear that by the screening of orbitals, the computational time required for the construction of the ACE operator has significantly decreased,  resulting in a net speed up in the force calculation per MD step;
%using the ACE operator; 
see $t^{\rm force}_{\textrm {s-ACE}}$ in Table~\ref{table1} for various values of $\rho_{\rm cut}$ and Figure~\ref{cost_ace}.
However, we have noticed that the number of SCF cycles required for computing energy/forces ($N^{\rm SCF}$) increases with screening, due to poor initial guess of wavefunctions.
%obtained using extrapolation. 
%
Due to this reason, the overall speed-up in 
%force calculation per MD step (on average) 
generating 1~fs long trajectory (on average)
is not proportional to $n$ (See Table~\ref{table_32}). 
For a 32 water periodic system, we could achieve the maximum speed up (of $\sim$9X) using $\rho_{\rm cut}=2.5 \times 10^{-2}$ with {\bf MTS-15}. 
The performance was better compared to lower values of $\rho_{\rm cut}$ for the same value of $n$ (Figure~\ref{scaling}). 
As expected, {\bf MTS-15} performed better than {\bf MTS-7} and {\bf MTS-5}.  
Speed-ups of $\sim$11X and $\sim$13X were observed for 64 and 128 water molecules systems, respectively.
%
%{\color{blue} However, we find that high values of $\rho_{\rm cut}$ does not guarantee better speed-up for a given value of $n$.}
%
%As discussed earlier, this is because, 
%We observed that that the number of SCF cycles per force calculation could increase with higher $\rho_{\rm cut}$ values due to relatively poor initial guess of wavefunction.

\def\arraystretch{1.30}
\begin{table*}
%\centering
\caption{\label{table_32} 
Average computational time {for generating 1~fs trajectory}
%NN MD step 
($t_{\rm CPU}$) for periodic systems containing 32, 64 and 128 water molecules
using PBE0.
%
%Total number of processors used for each system are reported in parenthesis. 
%
The averages were calculated over $N_{\rm MD}$ number of MD steps.
The achieved speed-ups for various {\bf MTS-n} runs are compared with the {\bf VV} run.\cite{processor_hpc} 
%
%{\NNc {cite {CITE-NOTE-WITH-CPU-DETAILS}}} \NNc{TODO-1: Correct bib entry here. TODO-2: Correct $t_{\rm CPU}$ from ``per step'' to ``per fs''}
%
%Here $\rho_{\rm cut}$ is in a.u.
%
%All the calculations were performed using identical 120 processors. 
%The comparison of different types of force calculation time for {\bf VV} and {\bf MTS} simulations.
}
%\resizebox{\textwidth}{!}
{%
%\begin{center}
%\begin{ruledtabular}
\begin{tabular}{|c|c|c|c|c|c|c|c|c|c|}
\hline \hline
\multirow{2}{*}{System size} & \multirow{2}{*}{No. of CPU cores} & \multirow{2}{*}{Method} & \multirow{2}{*}{$N_{\rm MD}$} & \multicolumn{2}{c|}{$\rho_{\rm cut}=2.0\times10^{-3}$} & \multicolumn{2}{c|}{$\rho_{\rm cut}=1.0\times10^{-2}$} & \multicolumn{2}{c|}{$\rho_{\rm cut}=2.5\times10^{-2}$} \\ \cline{5-10} 
                        &        & &                  & $t_{\rm CPU}$ (s)     & speed-up     & $t_{\rm CPU}$ (s)      & speed-up     & $t_{\rm CPU}$ (s)      & speed-up      \\ \hline \hline
%
%\multicolumn{8}{|c|}{32 water (120 processors)} \\ \hline 
%
\multirow{4}{*}{32 water} & \multirow{4}{*}{120} & {\bf VV}                     & 630                      & 518      & 1.0          & 509      & 1.0          & 512       & 1.0           \\ \cline{3-10}
& & {\bf MTS-5}                   & 630                      & 127      & 4.1          & 129      & 4.0          & 136      & 3.8           \\ \cline{3-10}
& & {\bf MTS-7}                   & 630                      & 103      & 5.0         & 98      & 5.2          & 101      & 5.1           \\ \cline{3-10}
& &  {\bf MTS-15}                  & 630                      & 70     & 7.4          & 57     & 8.9          & 56      & 9.2           \\ \hline \hline
%
%\multicolumn{8}{|c|}{64 water (160 processors)} \\ \hline 
%
\multirow{4}{*}{64 water} & \multirow{4}{*}{160} & {\bf VV}                     & 105                      & 3962      & 1.0          & 3969  & 1.0          & 3976     & 1.0           \\ \cline{3-10}
& & {\bf MTS-5}                   & 105                      & 856      & 4.6          & 936  & 4.2          & 1059     & 3.8           \\ \cline{3-10}
& & {\bf MTS-7}                   & 105                      & 656     & 6.0          & 692  & 5.7          & 792     & 5.0           \\ \cline{3-10}
& & {\bf MTS-15}                  & 105                      & 396      & 10.0          & 371  & 10.7          & 402      & 9.9           \\ \hline \hline
%
%
%\multicolumn{8}{|c|}{128 water (200 processors)} \\ \hline 
%
\multirow{4}{*}{128 water} & \multirow{4}{*}{200} & {\bf VV}                     & 5                      & 32668     & 1.0          & 32783      & 1.0          & 32726     & 1.0           \\ \cline{3-10}
& & {\bf MTS-5}                   & 15                      & 6400     & 5.1          & 7607      & 4.3          & 8525      & 3.8           \\ \cline{3-10}
& & {\bf MTS-7}                   & 21                      & 4914    & 6.6          & 5562     & 5.9          & 6020      & 5.4           \\ \cline{3-10}
& & {\bf MTS-15}                  & 45                      & 2611      & 12.5          & 2749     & 11.9          & 2998       & 10.9           \\ \hline \hline
\end{tabular}%
%\end{ruledtabular}
%\end{center}
}
\end{table*}

\begin{table*}
\caption{\label{table1} 
The decomposition of the total computing time for force calculations for a periodic system containing 32 water molecules. 
Timings are in seconds when using identical 120 CPU cores. 
%
%Intel$^{\textregistered}$ Xeon$^{\textregistered}$ CPU E5-2630v4 (20 cores/node) nodes.}\cite{processor_meggie} 
%
%{\color {red} Intel$^{\textregistered}$ Xeon$^{\textregistered}$ CPU E5-2670v2 (20 cores/node)}\cite{processor_hpc}
%\\ [2ex]
Here, $t_{\textrm{s-ACE}}^{\rm Con}$ is the computing time for the construction of ${\mathbf V}_{\rm X}^{\textrm{s-ACE}}$, $t_{\textrm{s-ACE}}$ is the computing time per SCF cycle using ${\mathbf V}_{\rm X}^{\textrm{s-ACE}}$, $N^{\rm SCF}_{\textrm{s-ACE}}$ is the average number of SCF cycles during the computation of $\mathbf F^{\textrm{s-ACE}}$, $t_{\textrm{s-ACE}}^{\rm force}$ is the total computing time for the calculation of $\mathbf F^{\textrm{s-ACE}}$, $t_{\rm exact}$ is the computing time per SCF cycle using ${\mathbf V}_{\rm X}$, 
$N^{\rm SCF}_{\rm exact}$ is the average number of SCF cycles for $\mathbf F^{\textrm{exact}}$ force calculation,  $t_{\rm exact}^{\rm force}$ is the total computing time for the calculation of $\mathbf F^{\textrm{exact}}$,  $t_{\textrm{s-ACE}}^{\rm force} = t_{\textrm{s-ACE}}^{\rm Con} + N^{\rm SCF}_{\textrm{s-ACE}} ~ t_{\textrm{s-ACE}} $, and  $t_{\rm exact}^{\rm force} = N^{\rm SCF}_{\rm exact} ~ t_{\rm exact} $.
}
%\resizebox{\textwidth}{!}
{%
%\begin{center}
\begin{ruledtabular}
\begin{tabular}{|c|c|c|c|c|c|c|c|c|}
%\hline \hline 
Method & $\rho_{\rm cut}$ & $t_{\textrm{s-ACE}}^{\rm Con}$ & $t_{\textrm{s-ACE}}$ & $N^{\rm SCF}_{\textrm{s-ACE}}$ & $t_{\textrm{s-ACE}}^{\rm force}$ & $t_{\rm exact}$ & $N^{\rm SCF}_{\rm exact}$  & $t_{\rm exact}^{\rm force}$  \\ \hline
{\bf VV} & -- & -- & -- & -- & -- & 23.6 & 10.4 & 245.4  \\ \hline
{\bf MTS-15} & without screening  & 23.6 & 0.15 & 10.1 & 25.1 & 23.6 & 8.0  & 188.8   \\ \hline
{\bf MTS-15} & $2.0\times10^{-3}$ & 17.7 & 0.15 & 10.9 & 19.3 & 23.6 & 8.6  & 203.0   \\ \hline
{\bf MTS-15} & $1.0\times10^{-2}$ & 8.6  & 0.15 & 12.6 & 10.5 & 23.6 & 11.0 & 259.6   \\ \hline
{\bf MTS-15} & $2.5\times10^{-2}$ & 6.0  & 0.15 & 13.4 & 8.0  & 23.6 & 12.0 & 283.2   \\ %\hline \hline
\end{tabular}%
\end{ruledtabular}
%\end{center}
}
\end{table*}

%\clearpage

%
%
%\clearpage
%
%

%\clearpage

%\clearpage

%\clearpage

%%%%%%%%%%%%%%%%%%%%%%%%%%%%%%%%%%%%%

%
%\begin{figure}
%\begin{center}
%\includegraphics[scale=0.65]{fes.eps}% Here is how to import EPS art
%\end{center}
%\caption{\label{fes} Free energy surfaces computed using 
%(a) {PBE}}
% 
%\end{figure}
%
%
%\begin{figure}
%\begin{center}
%\includegraphics[scale=0.65]{fes_hfx.eps}% Here is how to import EPS art
%\end{center}
%\caption{\label{fes} Free energy surfaces computed using 
%(a) {PBE0}}
% 
%\end{figure}
%

\subsection{\label{sec:application}Application: Methyl Formate Hydrolysis}
\begin{figure}[h]
\begin{center}
\includegraphics[scale=0.85]{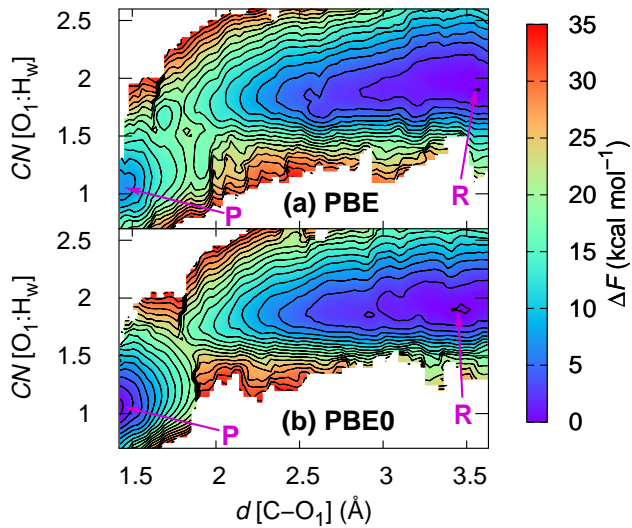}% Here is how to import EPS art
\end{center}
\caption{\label{fes} Free energy surfaces computed using 
(a) {PBE}
%(b) {\bf PBE+NSMD},
and 
(b) { PBE0}
density functionals for the {\bf R}$\rightarrow${\bf P} reaction are presented. Contour lines are drawn every 2 kcal mol$^{-1}$. 
%A representative snapshot of the {\bf TS} state
%is shown in (c). The indicated bond distances here are in {\AA}. Atom colors: red (O); blue (N); black (C); white (H).
}
\end{figure}
\begin{table}[h]
\caption{\label{table_fes} Free energy barriers ($\Delta F^\ddagger$) of 
the reaction {\bf R}$\rightarrow${\bf P} (Figure~\ref{mech}) 
using PBE and PBE0 functionals are compared
with the experimental estimate.}
\centering
%\begin{ruledtabular}
\begin{tabular}{|l|c|}
\hline \hline
Method & $\Delta F^\ddagger$ (kcal mol$^{-1}$)\\
\hline
{PBE} & 18.4  \\
{PBE0} & 20.8  \\
Experiment\cite{ester_exp2} & 28.8 \\
\hline \hline 
\end{tabular}
%\end{ruledtabular}
\end{table}

The hydrolysis of carboxylic esters is one of the well studied reaction in the field of chemistry, biochemistry and industrial chemistry.
Methyl formate hydrolysis serves as the simplest model for the hydrolysis of carboxylic esters and has been well studied experimentally\cite{ester_exp1,ester_exp2,ester_exp3,ester_exp4,ester_expt} and theoretically.\cite{ester_theo1,ester_theo2,ester_theo3,ester_theo4,ester_theo}
In our present work, we modelled the methyl formate hydrolysis reaction in neutral water.
This reaction happens through proton transfer between  solvent and the solute.
%
%Thus including explicit solvent molecules in the simulation is essential for an accurate estimate of the free energy barrier and mechanism.}
%
%The hydrolysis of carboxylic esters is considered here to demonstrate the applicability of our implementation in studying chemical reactions.
%is one of the extensively studied reaction in the field of chemistry, biochemistry and industrial chemistry.
%
%In particular, it serves as a model for the hydrolysis of peptide bonds in enzymes.
%
%In particular, methyl formate hydrolysis reaction has been well studied experimentally\cite{ester_exp1,ester_exp2,ester_exp3,ester_exp4,ester_expt} and theoretically.\cite{ester_theo1,ester_theo2,ester_theo3,ester_theo4,ester_theo} 
%
%Here, we investigate neutral hydrolysis of methyl formate.
%as an ideal model reaction to demonstrate the application of s-MTACE method.
%
Neutral hydrolysis of methyl formate leads to the formation of formic acid and methanol.
It follows a stepwise mechanism, where addition of a solvent water molecule to the carbonyl group results in a stable gem-diol intermediate (see Figure~\ref{mech}), followed by the decomposition of the intermediate to the final products.\cite{ester_theo,ester_theo4}
In our work, we focus only on the elementary step leading to the formation of the gem-diol intermediate.
%which is the rate determining step.\cite{ester_theo} 
%
%
The free energetics for this reaction step 
were computed using the WS-MTD technique.
%of formation of gem-diol intermediate from methyl formate in neutral aqueous medium.
%

For reweighting the WT-MTD bias potential, 
%we used the trajectories from a minimum time $t_{\rm min}$ to a maximum time $t_{\rm max}$.
%During the reconstruction of the free energy surface, we
%
%for all the umbrella windows, 
we used $t_{\rm min}=20$~ps and $t_{\rm max}= 35$~ps.
%for all the umbrella windows.
%
%Whereas, 
For the umbrella windows near the transition state region,
we took 
%we used $t_{\rm min}=15$~ps and different 
$t_{\rm max}$ as 46~(50)~ps for the PBE0 (PBE) simulations 
to obtain better statistics.
Here, $t_{\rm min}$ was chosen such that the free energy estimate is independent of the initial configuration.
%free energy estimate the initial structure.
%umbrella windows adjacent to the transition state.
%from which the simulations were started.
%
%In our simulations,
%we took the initial structure for an equilibration run from the adjacent equilibrated umbrella window.
%
The time series plots of the CVs for some of these umbrella windows
%$CN[{\mathrm {O_1:H_w}}]$ CV with simulation time for the umbrella windows near the transition state region 
are shown in Appendix~F.
%the Supporting Information, {Section~VI}.
%
%
%For few of the umbrellas, the system is stuck in wrong protonation state and could not explore the correct state during initial 15~ps.
%
%It remains stuck in the wrong state upto around 15~ps.
%
%During the simulation we observed that the system is stuck in the semi protonated state in the umbrellas of transition state regions.
%
%This is because of the dependency on the initial structure and system had the memory of that.
%
%In order to get history independent results, we took $t_{\rm min}=15$~ps when the system no longer retains the memory. 
%
%
%We performed WS-MTD simulations to investigate this particular step of the reaction.
%
The reconstructed free energy surfaces are shown in Figure~\ref{fes}. 
%
%The minimum energy pathway on the free energy surface was identified and free energy barrier was calculated along the pathway.
%on this surface was used to extract the mechanism and .
%
The convergence of free energy barriers as a function of $t_{\rm max}$ is reported in Appendix~G1.
%the Supporting Information, {Section~VII A}.
%
The converged free energy barriers and the experimental estimate are listed in Table~\ref{table_fes}.
From the free energy surfaces computed using both PBE and PBE0, it can be seen that the locations of the minimum corresponding to the reactant state {\bf R} and the intermediate {\bf P} are nearly the same. 
%
%deep minimum corresponding to the reactant state {\bf R} is nearly at the same location: $s_1 \in [2.9,3.5]$({\AA}), $s_2 \in [1.8,2.0]$ (unitless).
%
%The state {\bf P} is the tetrahedral intermediate, and the locations of {\bf P} are also nearly the same for both the functionals: $s_1 \sim 1.5$({\AA}), $s_2 \sim 1.0$ (unitless). 
%
%Similarly, the positions of the saddle point  ({\bf TS}) on the landscapes are also matching: $s_1 \sim 1.9$({\AA}), $s_2 \sim 1.3$ (unitless). 
%
%As expected, in the {\bf TS} structure, the attacking water molecule has dissociated one of its proton, 
%and a weak covalent bond between O$_1$ and the carbon atom is formed  (see Figure~\ref{fes3}(c)).
%
%This mechanism is in agreement with the previous work.
%
However, the topology of the free energy surfaces near the transition state region has some difference.
Additionally, the free energy barriers 
differ between the 
two functionals.
The converged free energy barrier computed using {PBE} is {18.4}~kcal/mol, while that using {PBE0} is {20.8}~kcal~mol$^{-1}$.  
The PBE0 free energy barrier is about {2.4}~kcal~mol$^{-1}$ higher than that of PBE, in line with the trends seen for the hydrolysis of formamide in water.\cite{JCP_sagar,sagar_JCC}
Notably, the free energy barrier computed from PBE0 is closer to the experimentally 
``estimated'' free energy barrier of 28.8~kcal~mol$^{-1}$ at 298~K\cite{ester_exp2}
for the reaction under neutral condition.
For simple amides, the free energy barrier for neutral hydrolysis was reported to be between 21.9 to 23.8~kcal~mol$^{-1}$.\cite{Callahan:jacs:05}
%THIS REFERENCE HAS PROBLEMS...ARE YOU SURE FREE ENERGY ESTIMATE IS IN THESE PAPERS? OR ARE YOU TAKING THE REF FROM OTHER PAPERS, WHICH CAN BE WRONG AS WELL...! PLEASE CAREFULLY CHECK}
%

%
 The PBE0 estimate of free energy barrier deviates substantially ($\sim$8~kcal~mol$^{-1}$) from the experimental data. 
This may be due to several reasons.
It is likely that under neutral aqueous conditions, the rate determining step is the subsequent elimination and not the formation of the gem-diol,\cite{chem_kin_2018_Allan} different to
what is observed under acidic condition.\cite{ester_exp1}
Finite size effects can also affect the results,
especially since the transition state configurations have an additional proton dissolved in water.
%, finite size effects are likely to influence the free energetics.
%
Clearly, a more detailed investigation is warranted in this direction to understand the deviation from the experimental data.
%require a more detailed investigation.
%
%
Nonetheless, the observed effect of hybrid functional on the free energy barrier is the most significant result of this study.
%
%For the same reaction, theoretical studies by Gunaydin {\it et al.}\cite{ester_theo4} and Tolosa Arroyo {\it et al.}\cite{ester_theo} reported a free energy barrier of 23.8 kcal/mol and 28.17 kcal/mol, respectively. 
%
We also note in passing that the previous computational studies have reported a range of free energy barriers for this reaction at 298~K.
Gunaydin {\it et al.}\cite{ester_theo4} reported that autoionization of water is the rate-determining step and the free energy barrier was predicted to be 23.8~kcal~mol$^{-1}$ (which in turn was taken from experiment data).
%wherein AIMD metadynamics with BLYP (GGA) functional was used.
%
%In their study, the reaction was initialized by autoionization of water for which the experimental free energy barrier is about 23.8 kcal~mol$^{-1}$.
%
%The subsequent hydrolysis step was found to have a free energy barrier of only {XX~kcal~mol$^{-1}$}.
%
Arroyo {\it et al.}\cite{ester_theo} reported a free energy barrier of 28.17 kcal~mol$^{-1}$ based on their MD simulations using empirical potentials.
%

%These MD simulations were performed with a solute-solvent potential that is derived from the MP2 level of theory. 
%
%with a discrete solvation model 
%
%The reaction coordinate used is the difference in the interaction energy of a given set of solvent molecules in the presence of the reactant and transition state structures.
%
%The free energy curves were constructed for both the transition state and the reactant state with solvent fluctuation as the reaction coordinate.
%
%
%Finally, the free energy barrier was computed from the locations of the minima on these curves.
%
%constructed along the reaction coordinate of the reactant and the product.
%It was concluded in their study that the activation barrier depends significantly on factors, like, chosen mechanism, level of theory used to model solute-solvent interaction
%
%The differences in the estimated free energy barrier may be due to the fact that different procedures were followed in these studies as compared to the present work.
%
%\NNc{$\Rightarrow$This part has to be simplified. Only these two works on ester? we can also give a comparison to amide/peptide hydrolysis data.}

{

\subsection{\label{sec:application1}Application: Proton Transfer Reactions within the Active Site Residues of Class-C $\beta$-Lactamase}
 $\beta$--Lactam antibiotics are commonly prescribed against bacterial infections. 
 They inhibit the bacterial cell--wall synthesizing enzymes known as penicillin binding proteins (PBPs).\cite{KleinE3463,Spencer:JMB:2019}
However, the clinical efficacy of these life saving drugs is progressively deteriorating due to the emergence of drug--resistance in bacteria, primarily associated with the expression of $\beta$--lactamases.\cite{Spencer:JMB:2019,Bush2010_2,Bebrone2010} 
These enzymes hydrolyze $\beta$-lactam antibiotics in an efficient manner, preventing the drug molecules to react with PBPs.
In the past, our group has contributed to the understanding
of the mechanism of hydrolysis of different classes of $\beta$--lactamases.\cite{Ravi_2012_JPCB,Ravi_2016_JPCB,Ravi_2013_JACS,Shalini_AZT,Chandan:ChemEurJ:2020,Chandan:pccp:2017} 
Here, we focus on class--C serine $\beta$--lactamases that is majorly responsible for hospital acquired infections.
It was found that the protonation states of the active site residues Lys$_{67}$ and Tyr$_{150}$ of class-C $\beta$-lactamase  play a crucial role in determining the  hydrolysis mechanism.\cite{Ravi_2012_JPCB,Ravi_2016_JPCB,Ravi_2013_JACS}
In the Michaelis complex of cephaloathin and the enzyme, two protonation states are possible: {$\bf K^+Y^-$} and {$\bf KY$}. 
In {$\bf K^+Y^-$}, Lys$_{67}$ is in the protonated form, while Tyr$_{150}$ is in its deprotonated form.
On the other hand, both Lys$_{67}$ and Tyr$_{150}$ are in their neutral forms in the {$\bf KY$} state (Figure~\ref{prot_mech}(a)).
Depending on the protonation state, the general base that activates Ser$_{64}$ could differ, rendering different acylation mechanisms. 

%Class-C $\beta-$lactamase active site has three crucial residues whose protonation states can critically determine the 
%structure and the reactivity: Lys$_{67}$, Tyr$_{150}$, and Lys$_{315}$.\cite{Ravi_2012_JPCB,Ravi_2013_JACS}
%
%The acylation and deacylation steps of the hydrolysis process involve a series of proton transfer reactions.
%, and it is believed that for both of these steps one of the active site residues acts as a general base to activate Ser$_{64}$ and the hydrolytic water, respectively.
%
%As a result, the relative stability of the protonation states of these residues becomes important in deciding the reactivity of the enzymes.
%
In the previous study,\cite{Ravi_2012_JPCB} the free energy barrier for the proton transfer between Lys$_{67}$ and Tyr$_{150}$ ({\bf K$^+$Y$^-\rightarrow$ KY}) was found to be small ($\sim1$~kcal/mol) at the level of PBE.
%
%Deprotonation of Lys67 is an important step to take place during the acylation reaction.\cite{ravi_JACS}
%
%The observation that the free energy barriers for proton transfer between Lys67 and Tyr150 is small could be an artifact of GGA density functionals used.\cite{Ravi_2012_JPCB}
%
It was found that, Lys$_{67}$ is hydrogen bonded to Ser$_{64}$ in the {\bf KY}  state and the former residue acts as the general base.
Here, we revisited the same problem and computed the free energy barrier separating the {\bf KY} and {\bf K$^+$Y$^-$} states.
It is known that PBE can underestimate the proton transfer barriers, and thus studying the above problem using a higher level of theory, especially including the contributions of HF exchange, is crucial.\cite{Adamo:2012}
We computed free energies using  QM/MM MD with PBE and PBE0  density functionals, and compared their performance.
%

%To make accurate predictions of free energies and to utilize the parallel computing environment in an efficient manner, we used the umbrella sampling technique\cite{US_method}.
%method\cite{shalini_TASS,shalini_WIRE} developed by us. 
%
%Hybrid DFT calculations were accelerated by using the r-RESPA+ACE method.\cite{JCP_2019_sagar}
%
%Both these techniques are implemented in the latest version of the CPMD program,\cite{cpmd} which will be used for our calculations.
%

%{\color {magenta} For the reweighting of umbrella windows, we used $t_{\rm min}=15$~ps and $t_{\rm max}=35$~ps for the PBE runs and $t_{\rm min}=5$~ps and $t_{\rm max}=25$~ps for the PBE0 runs.}
%
Umbrella sampling simulations were carried out and the reconstructed free energy surface along the coordinate $CN\mathrm{[Tyr_{150}O_\eta:Lys_{67}H_{\zeta}}]$ 
%{\color {red} (as defined in Equation~(\ref{CN}))} 
using the two density functionals are shown in Figure~\ref{prot_fes}.
The convergence of the free energy barriers as a function of simulation length is reported in Appendix~G2.
%the Supporting Information, Section~VII B.
%
Also, the error in the computed free energy was calculated using the method given by Hummer et al.\cite{Hummer:error}
%and has been reported in the Supporting Information.
%
For both the functionals, the location of the reactant minima ({\bf K$^+$Y$^-$}) is around $CN \in \left [0.3,0.4\right]$ (unitless).
Whereas, the product minima ({\bf KY}) is located near $CN \in \left [0.8,0.9\right]$ (unitless).
%$CN=0.80$ for PBE functional and close to $CN=0.90$ for PBE0 functional.
%
However, we can notice 
%that there are 
qualitative and quantitative differences in  free energy profiles. 
At the level of PBE,
%density functional, 
the free energy barrier for the proton transfer in forward direction ({\bf K$^+$Y$^-\rightarrow$ KY}) is $1.7$ kcal~mol$^{-1}$, 
and the barrier for the proton transfer in the reverse direction ({\bf KY $\rightarrow$ K$^+$Y$^-$}) is $0.9$ kcal~mol$^{-1}$.
On the other hand, at the PBE0 level, the barrier for the forward proton transfer is $1.2$ kcal~mol$^{-1}$, and the barrier for the reverse proton transfer  is $2.0$ kcal~mol$^{-1}$.
%
%The computed free energy barriers clearly indicate that both of the states are thermally accessible at room temperature.
%
The error in the free energy estimates is less than 0.3~kcal~mol$^{-1}$. 
Notably, {\bf K$^+$Y$^-$} is predicted to be the most stable state with PBE %calculation,
whereas, {\bf KY} is the most stable state with PBE0.
In the earlier study\cite{Ravi_2012_JPCB}  using PBE and  QM/MM metadynamics simulation, it was reported that both of the states are equally stable with only $\sim 1$ kcal~mol$^{-1}$ barrier for their inter conversion
%between the two protonation states 
(in the presence of the substrate). 
This is consistent with the results of the present work using PBE.
Although, at the PBE0 level, the most stable state is found to be different, the free energy difference between the two protonation states 
%is only $\sim$1~kcal~mol$^{-1}$  
and the free energy barriers for proton transfer reactions are
%only $\sim 1-2$~kcal~mol$^{-1}$. 
small.
This implies that quick proton transfer between the Tyr$_{150}$ and Lys$_{67}$ 
residues can take place at ambient temperature.
Therefore we conclude that class-C $\beta-$lactamase catalyzed  acylation mechanism of the hydrolysis of cephalothin, wherein the deprotonated form of Lys$_{67}$ activates Ser$_{64}$,\cite{Ravi_2013_JACS} 
is justified at the PBE0 level.
%But nothing can clearly be said about the relative stability of the states.
%
%\NNc{However, it is difficult to say anything conclusive about the kinetic and thermodynamic stabilities of the different protonation states, as the differences in the free energy barriers are very small (less than $1$ kcal/mol). 
%
%Most likely, both of the states will be equally accessible at room temperature, which will lead to rapid proton hopping between the two active site residues.}
%

%
\begin{figure}[h]
\begin{center}
\includegraphics[scale=0.7]{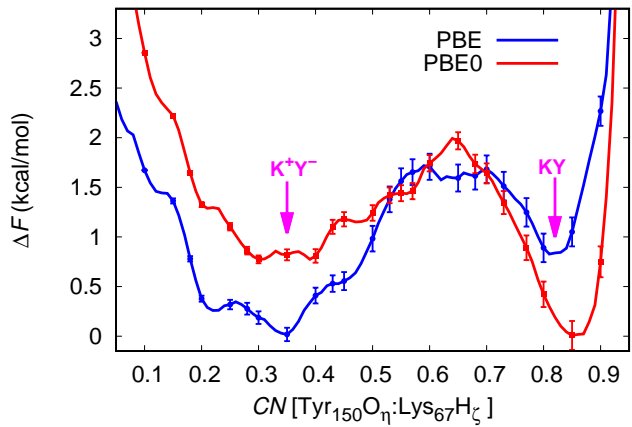}% Here is how to import EPS art
\end{center}
\caption{\label{prot_fes} Free energy as a function of $CN\mathrm{[Tyr_{150}O_\eta:Lys_{67}H_{\zeta}}]$ coordinate
computed for the proton transfer reaction ({\bf K$^+$Y$^-\rightleftharpoons$} {\bf KY}) using umbrella sampling
simulations at the level of {PBE} (blue)
%(b) {\bf PBE+NSMD},
and {PBE0} (red)
density functionals.  
{Errors in the free energy estimates were computed following Ref.~\onlinecite{Hummer:error}. }
%A representative snapshot of the {\bf TS} state
%is shown in (c). The indicated bond distances here are in {\AA}. Atom colors: red (O); blue (N); black (C); white (H).
}
\end{figure}

}

%\clearpage

\section{\label{sec:concl}Conclusions}

{
In summary, we presented a new scheme named s-MTACE to perform efficient hybrid functional based AIMD simulations using PW basis set.
%
%This is an extension of the MTACE method which is presented in the previous chapter.
%This involves artificial splitting in 
%Here, the ionic forces were split artificially to fast (computationally cheap) and slow (computationally costly) components using a screened ACE operator approach.
%
%This allows us to employ the r-RESPA scheme to speed-up the calculations. 
In particular, we have shown that screening the localized orbitals and computing ACE operator using the screened orbitals is an efficient way to speed up our earlier method.\cite{JCP_2019_sagar,sagar_JCC} 
Benchmark results show that dynamic and structural properties can be computed accurately using the s-MTACE method.
%with liquid water system show that stable and accurate MD trajectories can be obtained through this method. 
%
%Our profiling results show that the computational cost of the MTACE method decreases substantially by the introduction of the SCDM localization method in constructing the ACE operator.
%
We achieved a computational speed-up of $\approx$ 10 for a periodic 32-water molecules system compared to the conventional implementation of the hybrid functional PW-DFT.
For a larger system with 128 water molecules, we have achieved a speed-up of about 13.
%with this s-MTACE approach as opposed to the MTACE method where the speed up remained same.
%
%{\color {magenta} The {\tt CPMD} code wherein the s-MTACE method has been implemented will be made available on request before its official release.}

Using our implementation, we show that it is possible to carry out computationally demanding free energy calculations at the level of hybrid density functionals.
We demonstrated this by performing free energy calculations in two systems involving proton transfer reactions.
%
%
%NNR2
%Finally, .
%
First, our implementation was combined with the WS-MTD method to study the methyl formate hydrolysis in neutral aqueous solution.
The two dimensional free energy surface for this reaction was computed with PBE and PBE0 functionals.
%
%We found that the topologies of the free energy surfaces are different with these two different levels of functionals. 
%
We observed that the free energy barrier computed using PBE 
is 2.4 kcal/mol lower than PBE0, similar to our observations in earlier studies for a different system.\cite{JCP_sagar,sagar_JCC}
%, while the latter
}
By employing our method within the framework of  QM/MM, we investigated the proton transfer reaction between the active site residues in the Michaelis complex of class-C $\beta-$lactamase and cephalothin antibiotic.
We found that the stability of protonation states and the free energy barriers for proton transfer are altered while switching from PBE to PBE0.  

The general observation of underestimation of proton transfer barriers by PBE compared to PBE0 
is consistent with the detailed benchmarking studies by Adamo and co-workers\cite{Adamo:2012} 
through structural optimizations of several gas-phase reactions.
%NNR2
%By combining our implementation with well-sliced metadynamics,  it is now affordable to 
%
{
As this work enables us to perform free energy calculations at the level of hybrid functionals within AIMD simulations in an efficient manner, we hope that the presented method and the results obtained from our calculations are  of great importance to scientists working in
the area of computational catalysis.
%
%To our knowledge, this is the first work to do 
%a benchmark study on the influence of free energetics of chemical reactions involving proton transfers in enzymes at the two levels
%of the density functionals.
%
}

%%%%%%%%%%%%%%%%%%%%%%%%%%%%%%%%%%%%%%%%%%%%%%%%%%%%%%%%%%%%%%%%%%%%%%%%%%%%%%%

\begin{acknowledgments}

The authors thank Tobias Klöffel and Bernd Meyer for helpful discussions.
Financial support from the Science and Engineering Research Board (India)
under the MATRICS scheme (Ref. No. MTR/2019/000359) and from the German
Research Foundation (DFG) through Research Unit FOR 1878 (funCOS) and
Collaborative Research Center SFB 953 (project number 182849149) is
gratefully acknowledged.
SM and VT thank the University Grant Commission (UGC), India, and IITK for
their Ph.D. fellowships.
Computational resources were provided by the HPC facility (HPC2013)
at IITK, the Erlangen Regional Computing Center (RRZE) at FAU and
SuperMUC-NG (project pn98fa) at Leibniz Supercomputing Centre (LRZ).

%TODO cite Tobis paper \cite{KLOFFEL2021}
%Authors acknowledge the financial support from the Science and Engineering Research Board (India) under the
%MATRICS scheme (Ref. No. MTR/2019/000359).
%
%
%Calculations of Section~(\ref{bench_calc}) were carried out using the HPC facility (HPC2013) at the Indian Institute of Technology Kanpur (IITK).
%for the computational resources.
%
%Computations of Section~(\ref{sec:application}) were performed using the Meggie Compute-Cluster at the Erlangen Regional Computing Center (RRZE), Friedrich-Alexander-Universität Erlangen-Nürnberg (FAU).
%
%Computational resources for the Section~(\ref{sec:application1}) calculations were provided by the SuperMUC-NG (project pn98fa) at Leibniz Supercomputing Centre (LRZ). 
%through the project {pn98fa}.
%
%SM and VT thanks the University Grant Commission (UGC), India, and IIT Kanpur for their Ph.D. fellowships.
%
%VT thanks IIT Kanpur for her Ph.D. fellowship.
%
%SM acknowledges Prof. Bernd Meyer (FAU) 
%and DFG 
%for supporting his travel to
%to join his research group at the
%the Computer Chemistry Center (CCC), FAU. 
%as visiting research scholar.
%
%SM thanks {\color {red} DFG} for the financial support of his travel and stay at CCC through project {\color {red} XXX}.
%MATRICS
%Meggie
%Bernd
%DFG? travel support and stay 
%FAU
\end{acknowledgments}

%%%%%%%%%%%%%%%%%%%%%%%%%%%%%%%%%%%%%%%%APPENDIX%%%%%%%%%%%%%%%%%%%%%%%%%%%%%%%%%%%%%%%%%%%%%
\appendix
\renewcommand\thefigure{\thesection.\arabic{figure}}
\renewcommand\thetable{\thesection.\arabic{table}}
%%%%%%%%%%%%%%%%%%%%%%%%%%%%%%%%%%%%%%%%%%%%%%%%%%%%%%%%%%%%%%%%%%%%%%%%%%%%%%%%%%%%%%%%%%%%%
%
%

\section{Construction of the Adaptively Compressed Exchange Operator}
%
%The high computational cost of applying the ${\mathbf V}_{\rm X}$ operator on each occupied orbitals at each iteration step makes KS-DFT with hybrid functionals too expensive.
%
%In order to reduce the computational cost,
%
%Recently, Lin Lin developed the ACE operator formulation,\cite{ACE_Lin,ACE_Lin_1} to reduce the computational cost of such calculations.
%hybrid functional based DFT calculations.
%
According to the ACE operator formalism, the ACE operator (${\mathbf V}_{\rm X}^{\rm ACE}$) can be constructed through the following series of simple linear algebra operations.
%using a low rank decomposition.
%
%The ${\mathbf V}_{\rm X}^{\rm ACE}$  operator can be computed through a series of simpler linear algebra operations.
%
%Suppose we have a set of KS orbitals $\{|\phi_{j} \rangle\}$, 
First, the exact exchange operator (${\mathbf V}_{\rm X}$) has to be applied on the set of KS orbitals $\{|\psi_{i} \rangle\}$ as 
\begin{equation}
|W_{i}\rangle={\mathbf V}_{\rm X}|\psi _{i}\rangle , \enspace ~~i=1,....,N_{\rm orb} \enspace . 
%|W_{i}\rangle={\mathbf V}_{\rm X}[\{|\phi_{j} \rangle\}]|\phi _{i}\rangle , \enspace ~~i=1,....,N_{\rm orb} \enspace .
\end{equation}
Now, ACE formalism defines ${\mathbf V}_{\rm X}^{\rm ACE}$ as
\begin{equation}
{\mathbf V}_{\rm X}^{\rm ACE}= \sum_{i,j}^{N_{\rm orb}}  | W_{i} \rangle B_{ij}  \langle W_{j} | \enspace.
\end{equation}
Here, ${\mathbf B}={\mathbf M}^{-1}$, 
%is a negative semidefinite matrix. 
%
and the matrix ${\mathbf M}$ has elements 
\begin{equation}
%M_{kl}= \left \langle \psi_k | W_l \right \rangle   \enspace.
M_{kl}= \left \langle \psi_k | {\mathbf V}_{\rm X}|\psi _{l} \right \rangle   \enspace.
\end{equation}
Now, the Cholesky factorization of $-{\mathbf M}$ gives 
\begin{equation}
 {\mathbf M}=-{\mathbf L}{\mathbf L}^T  \enspace,
\end{equation}
where, ${\mathbf L}$ is a lower triangular matrix.
Then ${\mathbf B}$ is computed as
\begin{equation}
 {\mathbf B}=-{\mathbf L}^{-T}{\mathbf L}^{-1}  \enspace. 
\end{equation}
Finally, the ${\mathbf V}_{\rm X}^{\rm ACE}$ operator can be rewritten as
\begin{equation}
{\mathbf V}_{\rm X}^{\rm ACE}   = - \sum_{k}^{N_{\rm orb}} |P_{k} \rangle  \langle P_k |   \enspace,
\end{equation}
with $\{|P_{k} \rangle\}$ as the columns of the matrix ${\mathbf P}$, which is defined 
%matrix ${\mathbf P}$ with ACE projection vectors $\{|P_{k} \rangle\}$ as columns is defined
as
\begin{equation}
  {\mathbf P}= {\mathbf W}{\mathbf L}^{-T} \enspace.
\end{equation}

\section{Details of the Umbrella Sampling Bias Parameters}
\setcounter{figure}{0}
\setcounter{table}{0}

\subsection{Methyl Formate Hydrolysis}
\def\arraystretch{1.3}

\begin{table}[H] %remove 1.67 PBE
{
 \caption{\label{umbrella1} {Umbrella sampling parameters: Here,
 $d_h^{(0)}$ is in {\AA} and $\kappa_h$ is  in Hartree Bohr$^{-2}$.}
 }
 }
 %\footnotesize
\begin{center}
 \begin{ruledtabular}
{%\footnotesize
 \begin{tabular}{|ccc|ccc|}  %{|@{\hspace{1cm}}ccc@{\hspace{2cm}}|ccc@{\hspace{1cm}}|}
 %\begin{footnotesize}
  $h$ & $d_h^{(0)}$ & $\kappa_h$ & $h$ & $d_h^{(0)}$ & $\kappa_h$ \\
  %\end{footnotesize}
 \hline
   1 & 1.51 & 0.4 & 19 & 2.21 & 0.4  \\
   2 & 1.60 & 0.2 & 20 & 2.25 & 0.4  \\
   3 & 1.63 & 0.4 & 21 & 2.30 & 0.2  \\
   4 & 1.65 & 0.8 & 22 & 2.39 & 0.4  \\
   5 & 1.67 & 0.8 & 23 & 2.43 & 0.4  \\
   6 & 1.70 & 0.8 & 24 & 2.50 & 0.2  \\
   7 & 1.73 & 0.8 & 25 & 2.60 & 0.2  \\
   8 & 1.76 & 0.8 & 26 & 2.70 & 0.2  \\
   9 & 1.81 & 0.8 & 27 & 2.80 & 0.2  \\
  10 & 1.83 & 0.8 & 28 & 2.90 & 0.2  \\
  11 & 1.86 & 0.8 & 29 & 3.00 & 0.2  \\
  12 & 1.90 & 0.8 & 30 & 3.10 & 0.2  \\
  13 & 1.93 & 0.8 & 31 & 3.20 & 0.2  \\
  14 & 1.96 & 0.8 & 32 & 3.30 & 0.2  \\
  15 & 1.99 & 0.4 & 33 & 3.40 & 0.2  \\
  16 & 2.05 & 0.4 & 34 & 3.50 & 0.2  \\
  17 & 2.12 & 0.4 & 35 & 3.60 & 0.2  \\
  18 & 2.17 & 0.4 & 36 & 3.70 & 0.2  \\
 \end{tabular}
 }
 \end{ruledtabular}
 \end{center}
 %}
 \end{table}
% }

%%%%%%%%%%%%%%%%%%%%%%%%%%%%%%%%%%%%%%%%%%%%%%%%%%%%%%%%%%%%%%%%%%%%%%%55

\subsection{Protonation State of Active Site Residues of Class-C $\beta$-Lactamase}
\def\arraystretch{1.4}
\begin{table}[H] %remove 1.67 PBE
{
 \caption{\label{umbrella2} {Umbrella sampling parameters: Here,
 $d_h^{(0)}$ is unitless and $\kappa_h$ is  in Hartree.}
 }
 }
 %\footnotesize
\begin{center}
 \begin{ruledtabular}
{%\footnotesize
 \begin{tabular}{|ccc|ccc|} %{|@{\hspace{1cm}}ccc@{\hspace{2cm}}|ccc@{\hspace{1cm}}|} %  {|ccc|}
 %\begin{footnotesize}
  $h$ & $d_h^{(0)}$ & $\kappa_h$ & $h$ & $d_h^{(0)}$ & $\kappa_h$ \\
  %\end{footnotesize}
 \hline
   1 & 0.10 & 2.0  & 13 & 0.53 & 1.2  \\
   2 & 0.15 & 1.5  & 14 & 0.55 & 1.2   \\
   3 & 0.18 & 2.0  & 15 & 0.57 & 1.5   \\
   4 & 0.20 & 1.5  & 16 & 0.60 & 1.5   \\
   5 & 0.25 & 1.5  & 17 & 0.65 & 1.5   \\
   6 & 0.28 & 1.5  & 18 & 0.68 & 1.5   \\
   7 & 0.30 & 1.5  & 19 & 0.70 & 1.2   \\
   8 & 0.35 & 1.5  & 20 & 0.73 & 1.2   \\
   9 & 0.40 & 1.5  & 21 & 0.77 & 1.2   \\
  10 & 0.43 & 1.5  & 22 & 0.80 & 1.2   \\
  11 & 0.45 & 1.5  & 23 & 0.85 & 1.2  \\
  12 & 0.50 & 1.5  & 24 & 0.90 & 1.2  \\

 \end{tabular}
 }
 \end{ruledtabular}
 \end{center}
 %}
 \end{table}
% }

\section{Comparison of the Drift in Total Energy for VV and MTS Simulations in the NVE Ensemble}
\setcounter{figure}{0}
\setcounter{table}{0}
%\begin{center}
%
\def\arraystretch{0.95}
\begin{table}[h]
\caption{\label{table_drift_nve} The drift in total energy and $\log_{10}(\Delta E)$ for various {\bf VV} and {\bf MTS-n} simulations in $NVE$ ensemble.
%
%Drift was calculated as ${ \left |\left \langle  E(t) - E(0)  \right \rangle \right |}/ { {\rm number~of~atoms} / {\rm time}}$. Here, $E(t)$ is the total energy at any time $t$.
%
The results of {\bf MTS-n} simulations with various $\rho_{\rm cut}$ values are reported.
}
%\begin{center}
\begin{ruledtabular}
\begin{tabular}{ccccc}
%\hline \hline
Method & $\rho_{\rm cut}$  & Timestep  & Drift\footnote{Drift was calculated as ${ \left |\left \langle  E(t) - E(0)  \right \rangle \right |}/ { {\rm number~of~atoms} / {\rm time}}$. Here, $E(t)$ is the total energy at any time $t$.}  & $\log_{10}(\Delta E)$   \\
 &   &  (fs)  & (au/ps/atom)  &    \\
\hline
{\bf VV} & -- & 0.48 & 1.8$\times10^{-7}$ & -6.77  \\
{\bf VV} & -- & 0.96 & 1.6$\times10^{-6}$ & -6.14 \\
{\bf VV} & -- & 1.44 &  2.7$\times10^{-6}$ & -5.76  \\ \hline 
%
%   \multicolumn{4}{c}{$\rho_{\rm cut}=2.0\times10^{-3}$}  \\ 
   %\multicolumn{4}{|c|}{64 water (160 processors)} \\ \hline
\hline
{\bf MTS-3} & $2.0\times10^{-3}$ & 1.44 & 2.4$\times10^{-5}$ & -5.01  \\
{\bf MTS-5} & $2.0\times10^{-3}$ & 2.40 & 8.7$\times10^{-6}$  & -5.41 \\
{\bf MTS-7} & $2.0\times10^{-3}$ & 3.36 & 5.9$\times10^{-6}$  & -5.54 \\
{\bf MTS-15} & $2.0\times10^{-3}$ & 7.20 & 8.2$\times10^{-6}$  & -5.45  \\
{\bf MTS-30} & $2.0\times10^{-3}$ & 14.40 & 3.1$\times10^{-4}$ & -3.79  \\ \hline 
%
%\multicolumn{4}{c}{$\rho_{\rm cut}=1.0\times10^{-2}$} \\ 
\hline
{\bf MTS-3} & $1.0\times10^{-2}$ & 1.44 & 2.4$\times10^{-5}$ & -5.00  \\
{\bf MTS-5} & $1.0\times10^{-2}$ & 2.40 & 5.1$\times10^{-6}$  & -5.47 \\
{\bf MTS-7} & $1.0\times10^{-2}$ & 3.36 & 2.0$\times10^{-5}$  & -5.11 \\
{\bf MTS-15} & $1.0\times10^{-2}$ & 7.20 & 1.2$\times10^{-5}$  & -5.32  \\
{\bf MTS-30} & $1.0\times10^{-2}$ & 14.40 & 4.3$\times10^{-4}$ & -3.65  \\ \hline 
%
%\multicolumn{4}{c}{$\rho_{\rm cut}=2.5\times10^{-2}$} \\ 
\hline
{\bf MTS-3} & $2.5\times10^{-2}$ & 1.44 & 2.3$\times10^{-5}$ & -4.92  \\
{\bf MTS-5} & $2.5\times10^{-2}$ & 2.40 & 4.1$\times10^{-6}$  & -5.16 \\
{\bf MTS-7} & $2.5\times10^{-2}$ & 3.36 & 2.2$\times10^{-5}$  & -5.07 \\
{\bf MTS-15} & $2.5\times10^{-2}$ & 7.20 & 4.1$\times10^{-5}$  & -4.84  \\
{\bf MTS-30} & $2.5\times10^{-2}$ & 14.40 & 4.8$\times10^{-4}$ & -3.58  \\
%\hline \hline
%Experiment\cite{Formamide_Exp} & 21.2  \\
\end{tabular}
\end{ruledtabular}
%\end{center}
\end{table}

\section{%Comparison of Certain Properties using
Accuracy of VV and MTS Simulations in the NVT Ensemble}
\setcounter{figure}{0}
\setcounter{table}{0}
\def\arraystretch{1.0}
\begin{table}[h]
\caption{\label{table_drift_nvt} 
The simulation time length, average temperature ($\left < T \right >$) and drift in total energy for various {\bf VV} and {\bf MTS-n} (for various values of $\rho_{\rm cut}$) runs in $NVT$ ensemble.
%
%Drift was calculated as ${ \left |\left \langle  E(t) - E(0)  \right \rangle \right |}/ { {\rm number~of~atoms} / {\rm time}}$. Here, $E(t)$ is the total energy at any time $t$.
%
%The results of {\bf MTS-n} simulations for various  values are also reported.
%
%The comparison of different types of force calculation time for {\bf VV} and {\bf MTS} simulations.
}
\begin{center}
\begin{ruledtabular}
\begin{tabular}{cccccc}
%\hline \hline
 Method & $\rho_{\rm cut}$ & Timestep  & Length  & $\left \langle T \right \rangle$  & Drift\footnote{Drift was calculated as ${ \left |\left \langle  E(t) - E(0)  \right \rangle \right |}/ { {\rm number~of~atoms} / {\rm time}}$. Here, $E(t)$ is the total energy at any time $t$.}  \\
  &  & (fs)  & (ps)  &  (K)  & (au/ps/atom)  \\
 %
%  &  & (ps)  &  (K) &  (au/ps/atom) \\
\hline
  {\bf VV} & -- & 0.48 & 10 & 301 & $1.8\times 10^{-6}$  \\ \hline 
%
 %\multicolumn{5}{c}{$\rho_{\rm cut}=2.0\times10^{-3}$} \\
\hline
  {\bf MTS-5} & $2.0\times10^{-3}$ & 2.40 & 10 & 299 & $7.8\times 10^{-6}$ \\
  {\bf MTS-7} & $2.0\times10^{-3}$ & 3.36 & 10 & 299 & $1.0\times 10^{-5}$ \\
  {\bf MTS-15} & $2.0\times10^{-3}$ & 7.20 & 10 & 300 & $6.6\times 10^{-6}$ \\ \hline 
%
%\multicolumn{5}{c}{$\rho_{\rm cut}=1.0\times10^{-2}$} \\
\hline
  %{\bf VV} & 0.48 & 10 & 301 & $1.8\times 10^{-6}$ \\
  {\bf MTS-5} & $1.0\times10^{-2}$ & 2.40 & 10 & 300 & $7.6\times 10^{-6}$ \\
  {\bf MTS-7} & $1.0\times10^{-2}$ & 3.36 & 10 & 300 & $1.5\times 10^{-5}$ \\
  {\bf MTS-15} & $1.0\times10^{-2}$ & 7.20 & 10 & 300 & $8.0\times 10^{-6}$ \\  \hline 
%
%\multicolumn{5}{c}{$\rho_{\rm cut}=2.5\times10^{-2}$} \\
\hline
  %{\bf VV} & 0.48 & 10 & 301 & $1.8\times 10^{-6}$ \\
  {\bf MTS-5} & $2.5\times10^{-2}$ & 2.40 & 10 & 299 & $4.3\times 10^{-6}$ \\
  {\bf MTS-7} & $2.5\times10^{-2}$ & 3.36 & 10 & 299 & $1.7\times 10^{-5}$ \\
  {\bf MTS-15} & $2.5\times10^{-2}$ & 7.20 & 10 & 300 & $2.5\times 10^{-5}$ \\  
  %\hline \hline
% \hline
%  & {\bf VV} & 10 & 301 & $4.2\times 10^{-6}$ \\
%NVT & {\bf MTS-5} & 20 & 301 & $1.5\times 10^{-5}$ \\
% & {\bf MTS-15} & 20 & 300 & $1.1\times 10^{-5}$ \\
%Experiment\cite{Formamide_Exp} & 21.2  \\
\end{tabular}
\end{ruledtabular}
\end{center}
\end{table}

%%%%%%%%%%%%%%%%%%%%%%%%%%%%%%%%5
\section{Comparison of Structural and Dynamical Properties}
\setcounter{figure}{0}
\setcounter{table}{0}
\begin{figure}[H]
\centering\includegraphics[scale=0.38]{g_of_r_power_low.eps}
\caption{\label{gofr_low}
Radial distribution functions (RDFs) for bulk water 
(32 water system) simulation using {\bf VV}, {\bf MTS-5}, {\bf MTS-7} and {\bf MTS-15} trajectories at the level of PBE0: (a) O-O, (b) O-H,  and (c) H-H.
(d) Power spectrum of the same system computed from {\bf VV}, {\bf MTS-5}, {\bf MTS-7} and {\bf MTS-15} trajectories are 
shown.
Screening of the SCDMs was performed with $\rho_{\rm cut}=2.0\times10^{-3}$.
}
\end{figure}
\begin{figure}[H]
\centering\includegraphics[scale=0.38]{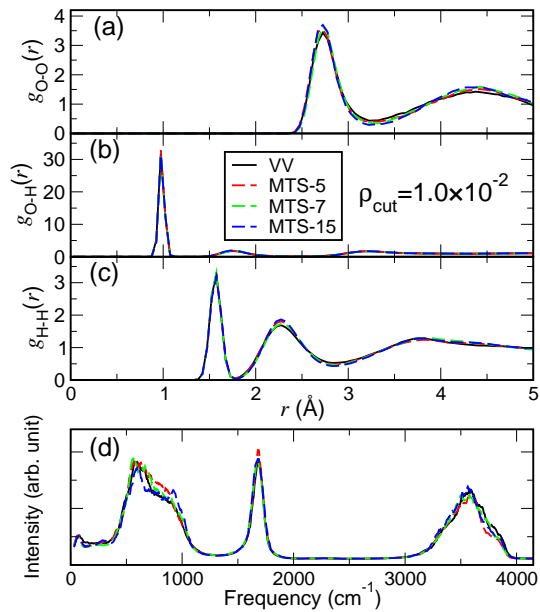}
\caption{\label{gofr_int}
Radial distribution functions (RDFs) for bulk water (32 water system) simulation using {\bf VV}, {\bf MTS-5}, {\bf MTS-7} and {\bf MTS-15} trajectories at the level of PBE0: (a) O-O, (b) O-H,  and (c) H-H.
(d) Power spectrum of the same system computed from {\bf VV}, {\bf MTS-5}, {\bf MTS-7} and {\bf MTS-15} trajectories are 
shown.
Screening of the SCDMs was performed with $\rho_{\rm cut}=1.0\times10^{-2}$.
}
\end{figure}
%

%

%%%%%%%%%%%%%%%%%%%%%%%%%%%%%%%%%%%%%%%%%%%%%%%%%%%%%%%%%%%%%%%%%%%%%%%55

\section{Time Series Plots of the CVs for Umbrella Windows Near the Transition State Region }
\setcounter{figure}{0}
\setcounter{table}{0}
\begin{figure}[H]
\begin{center}
\centering\includegraphics[scale=0.45]{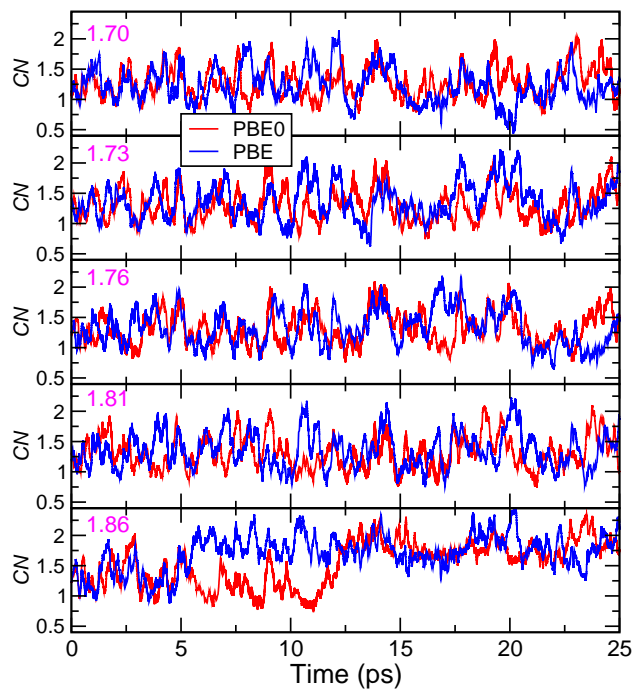}% Here is how to import EPS art
\end{center}
\caption{\label{cv_fluct1} Fluctuations in the $CN[{\mathrm {O_1:H_w}}]$ CV with simulation time for the umbrella windows near the transition state region (from 1.70 to 1.86~{\AA}) using PBE and PBE0 functionals. 
}
\end{figure}
\begin{figure}[H]
\begin{center}
\centering\includegraphics[scale=0.45]{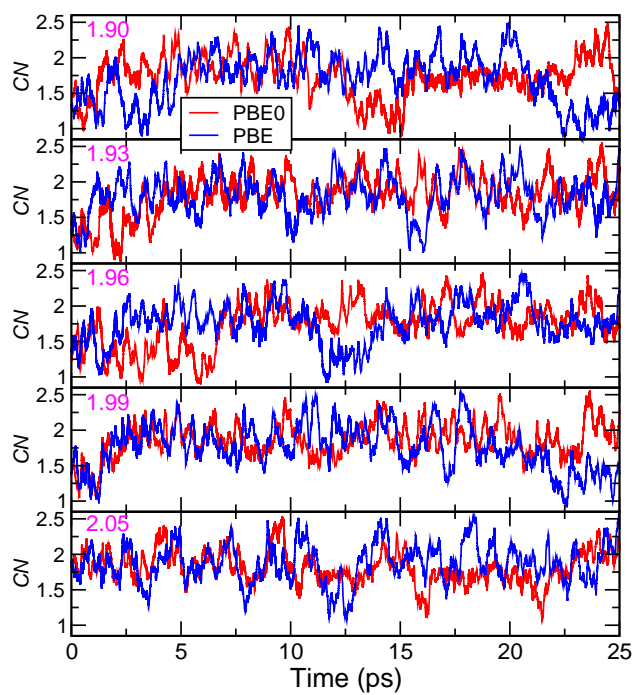}% Here is how to import EPS art
\end{center}
\caption{\label{cv_fluct2} Fluctuations in the $CN[{\mathrm {O_1:H_w}}]$ CV with simulation time for the umbrella windows near the transition state region (from 1.90 to 2.05~{\AA}) using PBE and PBE0 functionals. 
}
\end{figure}

\section{Convergence of the Free Energy Barriers}
\setcounter{figure}{0}
\setcounter{table}{0}
\subsection{Methyl Formate Hydrolysis}
\begin{figure}[H]
\begin{center}
\includegraphics[scale=0.7]{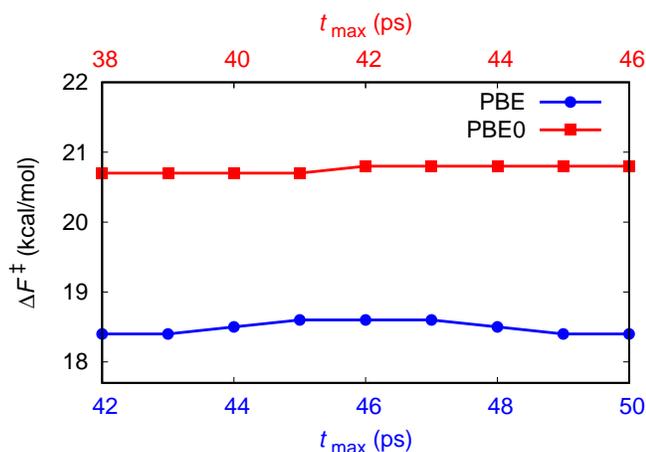}% Here is how to import EPS art
\end{center}
\caption{\label{fes_conv1} Convergence of the free energy barrier ($\Delta F^\ddagger$) for the reaction ${\bf R}\rightarrow{\bf P}$ with different $t_{\rm max}$ values using {PBE} and {PBE0} density functionals. Here, $t_{\rm min}=20$~ps was 
taken. 
}
\end{figure}

\subsection{Protonation State of Active Site Residues of Class-C $\beta$-Lactamase}
%
%\section{Convergence of the free energy barrier}
%
%{\color {red} TODO: add here (replace the figure with ur one)}
%
\begin{figure}[H]
\begin{center}
\includegraphics[scale=0.7]{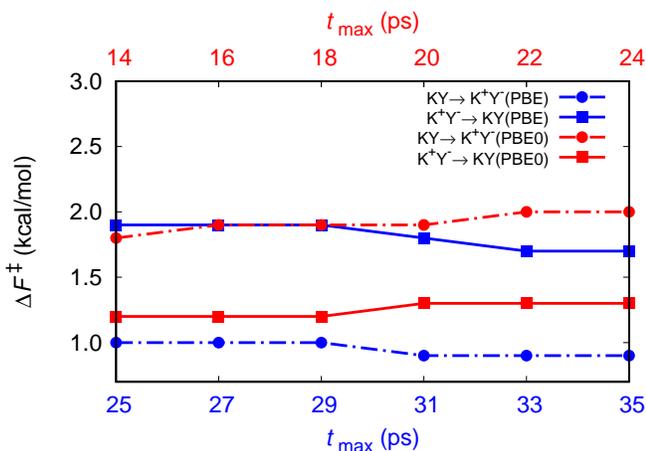}% Here is how to import EPS art
\end{center}
\caption{\label{fes_conv2} Convergence of the free energy barrier ($\Delta F^\ddagger$) with different $t_{\rm max}$ values for the reaction ${\bf KY}\rightarrow{\bf K^+Y^-}$ using PBE functional (blue circles) and PBE0 functional (red circles); and ${\bf K^+Y^-}\rightarrow{\bf KY}$ using PBE functional (blue squares) and PBE0 functional (red squares).
%using {PBE} density functional.
%
Here, $t_{\rm min}=15$~ps was taken for PBE functional; and $t_{\rm min}=4$~ps was taken for PBE0 functional. 
}
\end{figure}

\section*{References}

%%%%%%%%%%%%%%%%%%%%%%%%%%%%%%%%%%%%%%%%%%%%%%%%%%%%%%%%%%%%%%%%%%%%%%%%%%%%%%%%%%%%%
%\clearpage
%
%
%%%%%%%%%%%%%%%%%%%%%%%%%%%%%%%%%%%%%%%%%%%%%%%%%%%%%%%%%%%%%%%%%%%%%%%%%%%%%%%%%%%%%
%\nocite{*}
\bibliography{aipsamp}% Produces the bibliography via BibTeX.

\end{document}